\documentclass[%
 preprint,longbibliography,
 amsmath,amssymb,
 aps,prd
]{revtex4-1}

\newcommand\fft[2]{\frac{#1}{#2}}
\newcommand\ft[2]{{\textstyle\frac{#1}{#2}}}
\newcommand\nn{\nonumber}

\begin{document}

\preprint{MCTP-17-17}

\title{One-Loop Holographic Weyl Anomaly in Six Dimensions}

\author{James T. Liu}
\email{jimliu@umich.edu}
\affiliation{Michigan Center for Theoretical Physics, Randall Laboratory of Physics, The University of Michigan, Ann Arbor, MI 48109-1040}

\author{Brian McPeak}
\email{bmcpeak@umich.edu}
\affiliation{Michigan Center for Theoretical Physics, Randall Laboratory of Physics, The University of Michigan, Ann Arbor, MI 48109-1040}

\begin{abstract}

We compute $\mathcal O(1)$ corrections to the holographic Weyl anomaly for six-dimensional $\mathcal N=(1,0)$ and $(2,0)$ theories using the functional Schr\"odinger method that is conjectured to work for supersymmetric theories on Ricci-flat backgrounds. We show that these corrections vanish for long representations of the $\mathcal N=(1,0)$ theory, and we obtain an expression for $\delta(c-a)$ for short representations with maximum spin two. We also confirm that the one-loop corrections to the $\mathcal N=(2,0)$ M5-brane theory are equal and opposite to the anomaly for the free tensor multiplet. Finally, we discuss the possibility of extending the results to encompass multiplets with spins greater than two.

\end{abstract}

\maketitle


\section{Introduction}

The trace anomalies for conformal field theories provide an important set of quantities that characterize the theory.  In two dimensions, Cardy's formula \cite{Cardy:1986ie} demonstrates that the central charge $c$ is a reliable measure of the degrees of freedom.  Furthermore, its physical implication can be seen from the Zamolodchikov $c$-theorem \cite{Zamolodchikov:1986gt} which states that an effective $c$ function can be defined that is monotonically decreasing along renormalization group flows to the infrared.  While the picture is perhaps the clearest in two dimensions, recent work extending these results to higher dimensional CFTs has further emphasized the importance of trace anomalies in more general situations.

The AdS/CFT correspondence provides an ideal framework for investigating various anomalies, as they may often be reliably computed on both sides of the strong/weak coupling duality. Such calculations can provide a test of the AdS/CFT correspondence and can also yield additional insights on strongly coupled CFTs.  Here we focus on the conformal anomaly, which measures the change in the partition function that results from a Weyl scaling of the metric, $\delta g_{\mu\nu}=2\delta\sigma g_{\mu\nu}$. In particular, for a partition function given by
\begin{align}
    Z = \int \mathcal{D} \phi \exp{(-S[\phi])},
\end{align}
we define the anomaly $\mathcal{A}$ by
\begin{align}
    \delta\log Z = - \int d^d x \sqrt{\det g}\, \delta \sigma \mathcal A.
\end{align}
From the holographic point of view, the leading order Weyl anomaly can be obtained from the regularized classical action \cite{Henningson:1998gx}.  In the AdS$_5$/CFT$_4$ case, this gives the familiar result
\begin{equation}
    c=a=\frac{N^2}4\frac{\pi^3}{\mathrm{vol}(\Sigma^5)},
    \label{eq:aclead}
\end{equation}
where IIB supergravity has been compactified on AdS$_5\times\Sigma^5$.  Additional corrections to the leading order expression may arise from higher derivative modifications to the supergravity action as well as from quantum (i.e. loop) effects.

Holographically, the log-divergent part of the one-loop effective action provides an $\mathcal O(1)$ correction to the Weyl anomaly coefficients.  This was initially computed for the case of AdS$_5\times S^5$ in \cite{Bilal:1999ph,Mansfield:1999kk,Mansfield:2000zw,Mansfield:2002pa,Mansfield:2003gs}, where it was observed that the leading order result (\ref{eq:aclead}) is shifted according to $N^2\to N^2-1$, in agreement with expectations for $SU(N)$ gauge symmetry.  More recently, the one-loop computation in AdS$_5$ has been extended to holographic field theories with reduced or even no supersymmetry \cite{Ardehali:2013gra,Ardehali:2013xya,Ardehali:2013xla,Beccaria:2014xda}.

The one-loop holographic computation is essentially a sum over contributions from all states in the spectrum of single-trace operators.  Curiously, when arranged in terms of four-dimensional $\mathcal N=1$ superconformal multiplets, the contribution from long multiplets vanish identically.  As a result, only short representations contribute to the $\mathcal O(1)$ shift in $a$ and $c$.  This allows for a close connection between the central charges and the superconformal index which encodes knowledge of the shortened spectrum \cite{Ardehali:2014zba,Ardehali:2014esa} (see also \cite{DiPietro:2014bca}).

\subsection{The six-dimensional Weyl anomaly}

Here we wish to extend some of the holographic results for the Weyl anomaly coefficients to six dimensions.  In general, the anomaly takes the form
\begin{align}
 (4 \pi)^3 \mathcal A= (4 \pi)^3 \langle T \rangle = -a E_6 + (c_1 I_1 + c_2 I_2 + c_3 I_3) + D_\mu J^\mu,
\end{align}
where $E_6$ refers to the Euler density, $I_i$ are Weyl invariants and the final term is a non-universal total derivative.  At leading order, Einstein gravity on AdS$_7$ gives a relation of the form \cite{Henningson:1998gx}
\begin{equation}
    c_1=4c_2=-12c_3=96a\sim\mathcal O(N^3),
\end{equation}
which is the six-dimensional analog of (\ref{eq:aclead}).  The relation between the $c_i$ coefficients arises naturally in the holographic computation, and is further consistent with six-dimensional $(2,0)$ superconformal invariance.

The most extensively studied $(2,0)$ theory of relevance is that of $N$ coincident M5-branes, which is dual to supergravity on AdS$_7\times S^4$.  Here the conjectured expression for the central charges are \cite{Tseytlin:2000sf,Mansfield:2003bg,Beccaria:2014qea}
\begin{equation}
    a=-\frac{1}{288}(4N^3-\ft94N-\ft74),\qquad c=-\frac{1}{288}(4N^3-3N-1),
    \label{eq:M5ac}
\end{equation}
where $c_1=4c_2=-12c_3=96c$.  The $\mathcal O(N)$ terms arise from $R^4$ corrections \cite{Tseytlin:2000sf}, while the $\mathcal O(1)$ terms arise at one-loop \cite{Mansfield:2003bg,Beccaria:2014qea}.  The $\mathcal O(1)$ shift $\delta a=7/1152$ was computed in \cite{Beccaria:2014qea} by evaluating the one-loop partition function on global (Euclidean) AdS$_7$ with $S^6$ boundary.  However, the conjectured $\delta c=1/288$ has not yet been directly computed, as the most straightforward computation of one-loop determinants involve highly symmetric spaces with conformally flat boundaries.  In such cases, the Weyl invariants vanish, so no information is provided about the $c_i$ coefficients.

An alternative approach to the computation of $\delta a$ and $\delta c$ was developed in \cite{Mansfield:1999kk,Mansfield:2000zw,Mansfield:2002pa,Mansfield:2003gs} based on a functional Schr\"odinger approach.  In this approach, the contribution of each state to the $\mathcal O(1)$ shift in the Weyl anomaly takes the form
\begin{equation}
\delta\mathcal A=-\fft12\left(\Delta-\fft{d}2\right)b_d,
\label{eq:MN}
\end{equation}
where $\Delta$ is the conformal dimension and $b_d$ is the heat kernel coefficient for the corresponding AdS$_{d+1}$ field when restricted to the $d$-dimensional boundary.  In principle, since the six-dimensional $b_6$ coefficient may be computed on a general curved background, this allows for a full determination of not just the $a$ coefficient but the $c_i$'s as well.

It has been argued in \cite{Beccaria:2014xda}, however, that the expression (\ref{eq:MN}) cannot in general be valid, as the contribution for a single field should have a more complicated dependence on the conformal dimension $\Delta$.  This can be seen explicitly in comparison with the expression for $\delta a$ obtained directly from the one-loop determinant on global AdS.  Curiously, however, when (\ref{eq:MN}) is summed over the states of a complete supermultiplet, the resulting expression appears to be valid on Ricci-flat backgrounds as it passes all consistency checks and has the expected connection to the index \cite{Beccaria:2014xda,Ardehali:2014zba}.

In this paper, we use (\ref{eq:MN}) to compute the $\mathcal O(1)$ contribution to the holographic Weyl anomaly of $\mathcal N=(1,0)$ theories from maximum spin-2 multiplets in the bulk.  Since we consider Ricci-flat backgrounds, we only obtain information on $\delta(c-a)$, where $c$ is some linear combination of the Weyl coefficients (and reduces to the $c$ defined above in the $\mathcal N=(2,0)$ case).  This is similar to the AdS$_5$/CFT$_4$ case, where $b_4\sim\delta(c-a)R_{\mu\nu\rho\sigma}^2$ on Ricci-flat backgrounds.  As a consistency check, we find that $\delta(c-a)$ vanishes for long representations of $\mathcal N=(1,0)$ supersymmetry, as expected.

While we would ideally want an expression for the anomaly contribution from arbitrary higher-spin multiplets, this would require a better understanding of the heat kernel coefficients $b_6$ of higher-spin operators.  This was worked out in \cite{Christensen:1978gi,Christensen:1978md} for general spins in four dimensions.  However, a similar expression is lacking in six dimensions.  Nevertheless, knowledge of $\delta(c-a)$ for spins up to two is sufficient for computing the holographic Weyl anomaly for 11-dimensional supergravity on AdS$_7\times S^4$.  In this case, we sum the expression for $\delta(c-a)$ over the Kaluza-Klein spectrum and find the expected result $\delta(c-a)=-1/384$, in agreement with (\ref{eq:M5ac}) and the original computation of \cite{Mansfield:2003bg}.

\section{The $\mathcal O(1)$ contribution to the holographic Weyl anomaly} 

As indicated above, the six-dimensional Weyl anomaly may be parameterized as
\begin{equation}
    (4 \pi)^3 \mathcal A=-aE_6+(c_1I_1+c_2I_2+c_3I_3)+D_\mu J^\mu,
    \label{eq:Asix}
\end{equation}
where $E_6=\epsilon_6\epsilon_6RRR$ is the six-dimensional Euler density, and the Weyl invariants are given by
\begin{align}
    I_1&=C^a{}_{mn}{}^bC^m{}_{pq}{}^nC^p{}_{ab}{}^q,\nn\\
    I_2&=C^{ab}{}_{mn}C^{mn}{}_{pq}C^{pq}{}_{ab},\nn\\
    I_3&=C^{mnpq}\square C_{mnpq}+\cdots.
\end{align}
Superconformal invariance imposes additional constraints on the anomaly coefficients $\{a,c_i\}$.  In particular, $\mathcal N=(1,0)$ supersymmetry requires $c_1-2c_2+6c_3=0$, while $\mathcal N=(2,0)$ supersymmetry gives an additional constraint $c_1-4c_2=0$.  This suggests the parametrization
\begin{equation}
    c_1=96(c+c'+c''),\qquad c_2=24(c-c'+c''),\qquad c_3=-8(c+3c'-3c''),
\end{equation}
or equivalently
\begin{equation}
c=\fft{c_2-c_3}{32},\qquad c'=\fft{c_1-4c_2}{192},\qquad c''=\fft{c_1-2c_2+6c_3}{192}.
\label{eq:cc'c''}
\end{equation}
This is designed so that $c''$ vanishes for superconformal theories, and additionally $c'$ vanishes when there is extended $\mathcal N=(2,0)$ supersymmetry. The coefficient $c$ is chosen to allow for a quantity analogous to $c-a$ in four dimensions, as will become clear below.

The procedure we use to obtain the $\mathcal O(1)$ shift in the anomaly for $\mathcal N=(1,0)$ theories is to sum the expression (\ref{eq:MN}) over complete representations of the corresponding $OSp(8^*|2)$ supergroup.  However, we first start with states in the bosonic subgroup $OSp(8^*|2)\supset SO(2)\times SU(4)\times SU(2)_R$ labeled by $D(\Delta,j_1,j_2,j_3)$ along with $R$-symmetry representation $r$.  We thus have
\begin{equation}
    \delta\mathcal A(\mathrm{rep})=-\fft12\sum_{\mathrm{rep}}(\Delta-3)b_6(j_1,j_2,j_3).
    \label{eq:deltaA}
\end{equation}
In the following, we first work out the heat kernel coefficients $b_6(j_1,j_2,j_3)$ on a Ricci-flat background, and then perform the sum over complete supermultiplets with maximum spin two.

\subsection{Heat kernel coefficients}

For an operator $\Delta=-\nabla^2-E$ where $E$ is some endomorphism, the six-dimensional Seeley-DeWitt coefficient $b_{6}(\Delta)$ takes the form \cite{Gilkey:1975iq,Bastianelli:2000hi}
\begin{align}
    b_{6}(\Delta) &= \frac{1}{(4 \pi)^3 7!} \mathrm{Tr}\biggl[18A_1 + 17 A_2 - 2A_3 - 4 A_4 + 9 A_5 + 28 A_6 - 8 A_7 + 24 A_8 + 12 A_9 \nn\\
    &\quad+ \frac{35}{9} A_{10} -\frac{14}{3} A_{11} + \frac{14}{3} A_{12} - \frac{206}{9} A_{13} +\frac{64}{3} A_{14} -\frac{16}{3} A_{15} + \frac{44}{9} A_{16} +\frac{80}{9} A_{17}\nn \\
    &\quad+ 14 \Bigl( 8 V_1 + 2V_2 + 12V_3 -12 V_4 + 6V_5 -4V_6 +5V_7 +6V_8 + 60V_9 +30 V_{10} \nn\\
    &\kern2em+ 60 V_{11} + 30 V_{12} + 10 V_{13} + 4 V_{14} + 12 V_{15} + 30 V_{16} + 12V_{17} + 5 V_{18} -2 V_{19} + 2 V_{20}  \Bigl) \biggl].
\label{eq:b6coef}
\end{align}
Here the $A_a$'s form a basis of curvature invariants \cite{parker1987,Bastianelli:2000hi}, and the $V_a$'s are built from the endomorphism $E$ and the curvature $F_{ij}$ of the connection \cite{Bastianelli:2000hi}. In particular, while the coefficients of the $A_a$'s are universal, the $V_a$ terms are specific to the representation.

We follow the conventions spelled out in Appendix~A of \cite{Bastianelli:2000hi}, which also give explicit expressions for the $A_a$'s and $V_a$'s.  However, we are concerned with only the combinations which are non-vanishing on Ricci-flat backgrounds.  These are
\begin{equation}
A_5 = (\nabla_i R_{abcd})^2,\quad A_9 = R_{abcd} \nabla^2 R^{abcd},\quad
A_{16} = R_{ab}{}^{cd}R_{cd}{}^{ef}R_{ef}{}^{ab},\quad A_{17} = R_{aibj}R^{manb}R^i{}_m{}^j{}_n.
\end{equation}
The full list of $A_a$'s, and expressions for the $V_a$'s are given in Appendix~\ref{app:b6fields}.

The invariants $E_6$ and $I_1$, $I_2$, and $I_3$ may be written in terms of the basis $A_a$ functions. On a Ricci-flat background, they become
\begin{align}
    E_6 = 32 A_{16} - 64 A_{17} , \quad I_1 = - A_{17}  , \quad I_2 = A_{16}  , \quad    I_3 = 3A_5 + 6 A_9 + 2 A_{16} + 8 A_{17}.
    \label{eq:E6Ii}
\end{align}
As these quantities are not all independent, we will be unable to determine the individual central charges $\{a,c_i\}$ using only a Ricci-flat background.  Note that we may construct two combinations that are total derivatives
\begin{align}
    D_1&=\nabla_a(R_{mnij}\nabla_a R_{mnij})=A_5+A_9,\nn\\
    D_2&=2\nabla_a(R_{mnij}\nabla_m R_{anij})=-A_5+A_{16}+4A_{17}.
\end{align}
This allows us to rewrite (\ref{eq:E6Ii}) in terms of the two invariants $A_{16}$ and $A_{17}$
\begin{equation}
    E_6 =  32 A_{16} - 64 A_{17} , \quad I_1 = - A_{17}  , \quad I_2 = A_{16}  , \quad    I_3 = -A_{16} -4 A_{17}+6D_1-3D_2.
\end{equation}
On a Ricci-flat background, we have the relations $E_6= 32(2I_1+I_2)$ and $I_3=4I_1-I_2$ up to a total derivative.  As a result, the six-dimensional anomaly, (\ref{eq:Asix}), takes the form
\begin{equation}
   (4 \pi)^3 \mathcal A=32(c-a)A_{16}-64(c-a+3c'')A_{17}+D_\mu J^\mu,
    \label{eq:Aca}
\end{equation}
on Ricci-flat backgrounds.  The implication of this expression is that we will only be able to obtain information on the $\mathcal O(1)$ contribution to $c-a$ and to $c''$.  Since the latter must vanish for superconformal theories, it will serve as a consistency check of our approach.  This leaves us with a holographic determination of $\delta(c-a)$, which may be combined with the result of \cite{Beccaria:2014qea} for the $\delta a$ coefficient to extract both $\delta c$ and $\delta a$.  This, in principle, provides a complete determination of the $\mathcal O(1)$ shift in the holographic Weyl anomaly of $\mathcal N=(2,0)$ theories.  Unfortunately the additional anomaly coefficient $\delta c'$ for $\mathcal N=(1,0)$ theories cannot be determined in this manner on Ricci-flat backgrounds.

Ideally, we would like to have an expression for the heat kernel coefficient $b_6(\Delta)$ for fields transforming in an arbitrary $(j_1,j_2,j_3)$ representation of the six-dimensional $SU(4)$ Euclidean rotation group.  However, this requires understanding of arbitrary higher-spin Laplacians which currently eludes us.  There is also some potential ambiguity in relating `on-shell' states in AdS$_7$ to their corresponding boundary Laplacians in the functional Schrodinger approach of \cite{Mansfield:1999kk}.  We thus restrict to spins up to two.  The relevant $b_6$ coefficients evaluated on a Ricci-flat background are summarized in Table~\ref{tbl:b6coef}.  The coefficients for $\phi$, $\psi$, $A_\mu$ and $B_{\mu\nu}$ were computed in \cite{Bastianelli:2000hi}, while the remaining ones are worked out in Appendix~\ref{app:b6fields}.


\begin{table}[t]
\begin{center}
  \begin{tabular}{| c | c | c | c | c | c ||c|c|}
    \hline
    Field & $SU(4)$ Rep & $c_5$ & $c_9$ & $c_{16}$ & $c_{17}$&$\gamma_{16}$&$\gamma_{17}$ \\ \hline \hline
    $\phi$ & $(0,0,0)=\mathbf1$ & $9$ & $12$ & ${44}/{9}$ & ${80}/{9}$  &$17/9$&$-28/9$ \\ \hline
    $\psi$ & $(1,0,0)=\mathbf4$ & $-20$ & $-36$ & ${-202}/{9}$ & ${-436}/{9}$&$-58/9$&$140/9$  \\ \hline
    $A_{\mu}$ & $(0,1,0)=\mathbf6$ &$-58$ & $-96$ & ${-164}/{3}$ & ${-344}/{3}$&$-50/3$&$112/3$  \\ \hline
    $C^{+}_{\mu \nu \rho}$ & $(2,0,0)=\mathbf{10}$ & $174$ & $456$ & ${-5608}/{9}$ & ${26504}/{9}$&$-8146/9$&$16352/9$  \\ \hline
    $\Psi_{\mu}$ & $(1,1,0)=\mathbf{\overline{20}}$ & $292$ & $828$ & ${3526}/{9}$ & ${22012}/{9}$ &$-1298/9$&$2716/9$ \\ \hline
    $B_{\mu \nu}$ & $(1,0,1)=\mathbf{15}$ & $107$ & $348$ &${2992}/{3}$ & ${-1616}/{3}$ &$2269/3$&$-4508/3$ \\ \hline
    $G_{\mu \nu}$ & $(0,2,0)=\mathbf{20'}$ & $544$ & $1416$ & ${-1388}/{9}$ & ${49984}/{9}$ &$-9236/9$&$18592/9$ \\ \hline
  \end{tabular}
\end{center}
\caption{Heat kernel coefficients $(4\pi)^37!b_6=c_5A_5+c_9A_9+c_{16}A_{16}+c_{17}A_{17}$ for fields of spins up to two on a Ricci-flat background.  In the last two columns, we tabulate $\gamma_{16}$ and $\gamma_{17}$, where $(4\pi)^37!b_6=\gamma_{16}A_{16}+\gamma_{17}A_{17}+D_\mu J^\mu$.
\label{tbl:b6coef}}
\end{table}

\subsection{$\mathcal N=(1,0)$ Theory}

We now turn to the superconformal theories, starting with the $\mathcal N=(1,0)$ theory. We expect that the anomaly vanishes when summed over long representations, and we will see that this is indeed the case.  The $\mathcal N=(1,0)$ superconformal algebra is $OSp(8^*|2)$, with bosonic subgroup $SO(2,6)\times SU(2)_R$.  Here $SO(2,6)$ is either the isometry group of AdS$_7$ or the six-dimensional conformal group.  We label representations of $OSp(8^*|2)\supset SO(2,6)\times SU(2)_R\supset SO(2)\times SU(4)\times SU(2)_R$ by conformal dimension $\Delta$, $SU(4)$ Dynkin labels $(j_1,j_2,j_3)$ and $SU(2)_R$ Dynkin label $k$ (so that $SU(2)$ `spin' is given by $k/2$).

Unitary irreducible representations of the $\mathcal N=(1,0)$ theory have been studied and explicitly constructed in \cite{Minwalla:1997ka,Dobrev:2002dt,Bhattacharya:2008zy,Buican:2016hpb,Cordova:2016emh}.  The theory has one regular and three isolated short representations, given generically by
\begin{equation}
\begin{tabular}{ll}
$A[j_1,j_2,j_3;k]$:\qquad\hbox{}&$\Delta=\ft12(j_1+2j_2+3j_3)+2k+6$,\\
$B[j_1,j_2,0;k]$:&$\Delta=\ft12(j_1+2j_2)+2k+4$,\\
$C[j_1,0,0;k]$:&$\Delta=\ft12j_1+2k+2$,\\
$D[0,0,0;k]$:&$\Delta=2k$.
\end{tabular}
\label{eq:1,0short}
\end{equation}
For maximum spin two, however, we must restrict to $j_1=j_2=j_3=0$.  In this case, it is a simple exercise to perform the sum (\ref{eq:deltaA}) over the multiplet using the values of $\gamma_{16}$ and $\gamma_{17}$ given in Table~\ref{tbl:b6coef}.  Comparison with (\ref{eq:Aca}) then allows us to extract $\delta(c-a)$ and $\delta c''$.  The results are summarized in Table~\ref{tbl:(1,0)}.

\begin{table}[t]
\begin{tabular}{| l | c | c | c | c | c | c |}
 \hline
 &&$D[0,0,0;k]$&$C[0,0,0;k]$&$B[0,0,0;k]$&$A[0,0,0;k]$&$L[0,0,0;k]$\\
 Level & $SU(4)$ & $\Delta=2k$ & $\Delta = 2k + 2$ &$\Delta = 2k+4$ & $\Delta = 2k + 6$ & $\Delta > 2k+ 6$ \\ \hline
 $\Delta$ & $\mathbf1$ & $\scriptstyle k$ &$\scriptstyle k$ & $\scriptstyle k$ & $\scriptstyle k$ & $\scriptstyle k$ \\
 \hline
 $\Delta + \frac{1}{2}$ & $\mathbf4$ & $\scriptstyle k-1$ & $\scriptstyle k-1, k+1$ & $\scriptstyle k-1, k+1$ & $\scriptstyle k-1, k+1$ & $\scriptstyle k-1, k+1$ \\
 \hline
 $\Delta + 1$ & $\mathbf{10}$ &  & $\scriptstyle k$ & $\scriptstyle k$ & $\scriptstyle k$ & $\scriptstyle k$ \\
         & $\mathbf6$ & $\scriptstyle k-2$ & $\scriptstyle k-2, k$ & $\scriptstyle k-2, k, k+2$ & $\scriptstyle k-2, k, k+2$ & $\scriptstyle k-2, k, k+2$ \\
 \hline
 $\Delta + \frac{3}{2}$ & $\mathbf{\overline{20}}$ &  & $\scriptstyle k-1$ & $\scriptstyle k-1, k+1$ & $\scriptstyle k-1, k+1$ & $\scriptstyle k-1, k+1$ \\
         & $\mathbf{\overline4}$ & $\scriptstyle k-3$ & $\scriptstyle k-3, k-1$ & $\scriptstyle k-3, k-1, k+1$ & $\scriptstyle k-3, k-1, k+1, k+3$ & $\scriptstyle k-3, k-1, k+1, k+3$ \\
 \hline
 $\Delta + 2$ & $\mathbf{20'}$ & & & $\scriptstyle k$ & $\scriptstyle k$ & $\scriptstyle k$ \\
         & $\mathbf{15}$  & & $\scriptstyle k-2$ & $\scriptstyle k-2, k$ & $\scriptstyle k-2, k, k+2$ & $\scriptstyle k-2, k, k+2$ \\
         & $\mathbf1$   & $\scriptstyle k-4$ & $\scriptstyle k-4, k-2$ & $\scriptstyle k-4, k-2, k$ & $\scriptstyle k-4, k-2, k, k+2$ & $\scriptstyle k-4, k-2, k, k+2, k+4$ \\
         \hline
$\Delta + \frac{5}{2}$ & $\mathbf{20}$ &  &  & $\scriptstyle k-1$ & $\scriptstyle k-1, k+1$ & $\scriptstyle k-1, k+1$ \\
         & $\mathbf4$ & & $\scriptstyle k-3$ & $\scriptstyle k-3, k-1$ & $\scriptstyle k-3, k-1, k+1$ & $\scriptstyle k-3, k-1, k+1, k+3$ \\
         \hline
$\Delta + 3$ & $\mathbf{\overline{10}}$ &  &  &  &$\scriptstyle k$ & $\scriptstyle k$ \\
         & $\mathbf6$ &  &  & $\scriptstyle k-2$ & $\scriptstyle k-2, k$ & $\scriptstyle k-2, k, k+2$\\
 \hline
$\Delta + \frac{7}{2}$ & $\mathbf{\overline4}$ & & & & $\scriptstyle k-1$ & $\scriptstyle k-1, k+1$ \\
\hline
$\Delta + 4$ & $\mathbf1$ &  &  &  &  & $\scriptstyle k$\\
\hline
\hline
Anomaly&$2^5\cdot6!\delta(c-a)$ & $1$ &$57+180k$&$303+180k$& $-1$ & $0$  \\
&$\delta c''$     & $0$ & $0$ & $0$ & $0$ & $0$ \\        
\hline      
\end{tabular}
\caption{The $\mathcal N=(1,0)$ multiplets with maximum spin two, and corresponding holographic Weyl anomaly coefficients $\delta(c-a)$ and $\delta c''$.  Here $k$ is the $SU(2)_R$ Dynkin label (with spin~$=k/2$).  The shortening conditions correspond to those of (\ref{eq:1,0short}), while the last column is the maximum spin-two long representation.}
\label{tbl:(1,0)}
\end{table}

As a consistency check, we note that the anomaly coefficient $c''$ vanishes identically after summation over a complete multiplet.  This is a requirement of supersymmetry, but is not manifest from the individual $b_6$ coefficients in Table~\ref{tbl:b6coef}.  We also see that the anomaly vanishes for the long representation, which agrees with expectations from the AdS$_5$ case \cite{Ardehali:2013gra,Beccaria:2014xda}.  As for the non-vanishing contributions, note that $\delta(c-a)$ for the $A$ and $D$ type multiplets are equal and opposite. This must be the case, as $A[0,0,0;k]$ and $D[0,0,0;k+2]$ are ``mirror shorts" that sum to become a long multiplet.

Finally, recall that the $\mathcal N=(1,0)$ theory admits three independent anomaly coefficients, which we have parametrized as $a$, $c$ and $c'$.  Since we only consider Ricci-flat backgrounds, we have only been able to determine the difference $\delta(c-a)$.  This may be combined with the holographic $\delta a$ coefficient obtained in \cite{Beccaria:2014qea} to separate out the contributions to $\delta a$ and $\delta c$.  These results are presented in Table~\ref{tbl:10ac}.  However, we are unable to determine $\delta c'$ unless we can move away from Ricci-flat backgrounds.

\begin{table}[t]
\begin{tabular}{| c | c | c | c |}
    \hline
    Multiplet & $\Delta$ & $2^5\cdot6!\delta a$ & $2^5\cdot6!\delta(c-a)$\\
    \hline
    $L[0,0,0;k]$ & $>2k+6$ & $0$ & $0$  \\
    \hline
    $A[0,0,0;k]$ & $2k+6$ & $10\Delta^2(\Delta^2-2)+\fft{11}3$ & $-1$ \\
    \hline
    $B[0,0,0;k]$ & $2k+4$ & $-10(\Delta-\fft23)^2(3(\Delta-\fft23)^2-14)-\fft{530}9(\Delta-\fft23)-\fft{419}9 $ & $90(\Delta-\fft23)+3$\\
    \hline
    $C[0,0,0;k]$ & $2k+2$ & $10(\Delta-\fft43)^2(3(\Delta-\fft43)^2-14)-\fft{530}9(\Delta-\fft43)+\fft{419}9$ & $90(\Delta-\fft43)-3$ \\
    \hline
    $D[0,0,0;k]$ & $2k$ & $-10(\Delta-2)^2((\Delta-2)^2-2)-\fft{11}3$ & $1$ \\
    \hline
\end{tabular}

\caption{Contribution to the Weyl anomaly coefficients $\delta a$ and $\delta c$ from maximum spin two multiplets for the $\mathcal N=(1,0)$ theory. Here $c$ is related to the conventional anomaly coefficients $c_i$ according to (\ref{eq:cc'c''}).  The $\delta a$ coefficient is computed using the results of \cite{Beccaria:2014qea}.}
\label{tbl:10ac}
\end{table}

\subsection{$\mathcal N=(2,0)$ Theory}

We may perform the same computation for the $\mathcal N=(2,0)$ theory, noting however that only the 1/2-BPS multiplets have spins less than or equal to two.  In this case, the superconformal algebra decomposes as $OSp(8^*|4)\supset SO(2,6)\times Sp(4)_R\supset SO(2)\times SU(4)\times Sp(4)_R$.  The shortening conditions follow the same pattern as (\ref{eq:1,0short}), however with extended $R$-symmetry \cite{Minwalla:1997ka,Dobrev:2002dt,Bhattacharya:2008zy,Buican:2016hpb,Cordova:2016emh}
\begin{equation}
\begin{tabular}{ll}
$A[j_1,j_2,j_3;k_1,k_2]$:\qquad\hbox{}&$\Delta=\ft12(j_1+2j_2+3j_3)+2(k_1+k_2)+6$,\\
$B[j_1,j_2,0;k_1,k_2]$:&$\Delta=\ft12(j_1+2j_2)+2(k_1+k_2)+4$,\\
$C[j_1,0,0;k_1,k_2]$:&$\Delta=\ft12j_1+2(k_1+k_2)+2$,\\
$D[0,0,0;k_1,k_2]$:&$\Delta=2(k_1+k_2)$.
\end{tabular}
\end{equation}
Here $(k_1,k_2)$ are Dynkin labels for $Sp(4)$, with $(1,0)$ denoting the $\mathbf4$ and $(0,1)$ denoting the $\mathbf5$.  For maximum spin two, we restrict to the 1/2-BPS multiplets $D[0,0,0;0,k]$ with $\Delta=2k$.  (The case $k=1$ is the free tensor multiplet, while $k=2$ is the stress tensor multiplet.)

\begin{table}[t]
\begin{tabular}{| c | c | c | c | c |}
    \hline
    $\Delta$ & $SU(4)$ & $D[0,0,0;0,2]$ & $D[0,0,0;0,3]$ &  $D[0,0,0;0,k\ge4]$\\
    \hline
    \hline
    $2k$ & $\mathbf1$ & $(0,2)$ & $(0,3)$ & $(0, k)$ \\
    \hline
    $2k + \frac{1}{2}$ & $\mathbf4$ & $(1,1)$ & $(1,2)$ & $(1, k-1)$\\
    \hline
    $2k + 1$ & $\mathbf6$ & $(2,0)$ & $(2,1)$ & $(2, k-2)$\\
            & $\mathbf{10}$ & $(0,1)$ & $(0,2)$ & $(0, k-1)$\\
    \hline
    $2k + \frac{3}{2}$ & $\mathbf{\overline{20}}$ & $(1,0)$ & $(1,1)$ & $(1,k-2)$\\
                    & $\mathbf{\overline4}$ & & $(3,0)$ & $(3,k-3)$\\
    \hline
    $2k + 2$ & $\mathbf{20'}$ & $(0,0)$ & $(0,1)$ & $(0,k-2)$ \\
            & $\mathbf{15}$ & & $(2,0)$ & $(2,k-3)$\\
            & $\mathbf1$ & & & $(4,k-4)$\\
    \hline
    $2k+\frac{5}{2}$& $\mathbf{20}$ & & $(1,0)$ & $(1,k-3)$ \\
                & $\mathbf4$ & & & $(3, k-4)$\\
    \hline
    $2k+3$ & $\mathbf{\overline{10}}$ & & $(0,0)$ & $(0,k-3)$\\
        & $\mathbf6$ & & & $(2,k-4)$ \\
    \hline
    $2k+\frac{7}{2}$& $\mathbf{\overline4}$ & & & $(1,k-4)$\\
    \hline
    $2k+4$ & $\mathbf1$ & & & $(0,k-4)$\\
    \hline
    \hline
    Anomaly & $384\delta(c-a)$ & $13$ & $37$ & $6k(k-1)+1$\\ 
            &$\delta c''$&$0$&$0$&$0$ \\
    \hline
\end{tabular}

\caption{The $\mathcal N=(2,0)$ 1/2-BPS (maximum spin two) representation $D[0,0,0;0,k]$ and corresponding holographic Weyl anomaly coefficients $\delta(c-a)$ and $\delta c''$.  Entries are $Sp(4)_R$ representations specified by Dynkin labels $(k_1,k_2)$.}
\label{tbl:(2,0)}
\end{table}

The holographic computation of $\delta(c-a)$ and $\delta c''$ for the $D[0,0,0;0,k]$ multiplets are shown in Table~\ref{tbl:(2,0)}.  The case $k\ge4$ is generic, and we do not include $k=1$, which is a supersingleton and would not appear in a holographic computation.  The special case $k=3$ fits into the generic pattern, and in fact so does $k=2$, although it requires separate treatment because of the presence of massless modes.  For $k=2$, the states in $D[0,0,0;0,2]$ are
\begin{align}
    &D(4;0,0,0)_{\mathbf{14}}+D(4\ft12;1,0,0)_{\mathbf{16}}+D(5;0,1,0)_{\mathbf{10}}\nonumber\\
    &+D(5;2,0,0)_{\mathbf5}+D(5\ft12;1,1,0)_{\mathbf4}+D(6;0,2,0)_{\mathbf1},
\end{align}
where $D(\Delta;j_1,j_2,j_3)$ labels the $SO(2,6)$ representation and the subscript labels the $Sp(4)_R$ representation.  The massless vector, gravitino and graviton representations can be obtained from the corresponding massive representations by subtracting out null states according to
\begin{align}
    D(5;0,1,0)&=D(5+\epsilon;0,1,0)-D(6;0,0,0),\nn\\
    D(5\ft12;1,1,0)&=D(5\ft12+\epsilon;1,1,0)-D(6\ft12;1,0,0),\nn\\
    D(6;0,2,0)&=D(6+\epsilon;0,2,0)-D(7;0,1,0).
\end{align}
(Note that the three-form, $D(5;2,0,0)$, is massive, so no subtraction is required.) Taking these null states into account then gives the result $\delta(c-a)=13/384$ for $k=2$ shown in Table~\ref{tbl:(2,0)}.

Although $k=2$ and $k=3$ are special cases, the holographic anomaly coefficient $\delta(c-a)=(1/384)(6k(k-1)+1)$ is in fact universal.  Combining this with $\delta a=-(7/1152)(6k(k-1)+1)$ obtained in \cite{Beccaria:2014qea} then allows us to separate out the individual coefficients
\begin{equation}
    D[0,0,0;0,k\ge2]:\qquad\delta a=-\fft1{288}\cdot\fft74\left(6k(k-1)+1\right),\qquad\delta c=-\fft1{288}\left(6k(k-1)+1\right).
\end{equation}

As an application, consider the $\mathcal N=(2,0)$ theory obtained by compactifying 11-dimensional supergravity on AdS$_7\times S^4$.  The Kaluza-Klein spectrum is simply
\begin{equation}
\oplus_{k\ge2}D[0,0,0;0,k],
\end{equation}
where $k=2$ corresponds to the `massless' supergravity sector.
The anomaly coefficients $\delta a$ and $\delta c$ may be computed by summing over the Kaluza-Klein levels
\begin{equation}
     \delta a=-\fft1{288}\cdot\fft74\sum_{k=2}^\infty(6k(k-1)+1),\qquad
     \delta c=-\fft1{288}\sum_{k=2}^\infty(6k(k-1)+1).
\end{equation}
Following \cite{Beccaria:2014qea}, we regulate the sums using a hard cutoff. This amounts to setting $\sum_{k = 1}^{\infty} k^n = 0$ for any $n \geq0$. This implies $\sum_{k = 2}^\infty f(k)=-f(1)$, where $f(k)$ is polynomial in $k$.  As a result, the regulated anomaly for AdS$_7\times S^4$ is
\begin{equation}
     \delta a=\fft1{288}\cdot\fft74,\qquad\delta c=\fft1{288}.
\end{equation}
This is equal and opposite to the result for the conformal anomaly of the free tensor multiplet computed in \cite{Tseytlin:2000sf}, and agrees with the $\mathcal O(1)$ contributions in (\ref{eq:M5ac}).

\section{Discussion}

While the six-dimensional Weyl anomaly coefficients are conventionally parametrized as $a$, $c_1$, $c_2$ and $c_3$, we have found it convenient to use an alternate linear combination of the $c_i$'s given in (\ref{eq:cc'c''}).  Holographically, the leading order anomaly coefficients (assuming Einstein gravity in the bulk) satisfy the relation
\begin{equation}
    c=a,\qquad c'=0,\qquad c''=0.
\end{equation}
(While $c''=0$ must hold for superconformal theories, this holographic result is independent of whether the theory is supersymmetric or not.)
At the one-loop level, we have been able to extend this leading order result by computing $\delta(c-a)$ using the expression (\ref{eq:MN}) obtained through a functional Schr\"odinger method \cite{Mansfield:1999kk,Mansfield:2000zw,Mansfield:2002pa,Mansfield:2003gs}.

It is reasonable to question whether the use of (\ref{eq:MN}) is valid, as it disagrees with the direct computation of $\delta a$ performed in \cite{Beccaria:2014qea,Beccaria:2014xda}.  A quick way to see this is to note that $\delta a$ in Table~\ref{tbl:10ac} is a fourth order polynomial in $\Delta$, while the result of summing (\ref{eq:MN}) over a supermultiplet can be at most quadratic in $\Delta$.  (One power comes directly from (\ref{eq:MN}), while another can arise from the dimension of the shortened representation.)  If $\delta(c-a)$ was expected to be cubic or higher in $\Delta$, then our result, as shown in the last column of Table~\ref{tbl:10ac}, cannot possibly be correct. However, we now demonstrate that $c-a$ can be at most linear in $\Delta$, which is consistent with application of (\ref{eq:MN}).

To see this, recall that, in superconformal field theories, the stress tensor is contained in a multiplet of currents, so that there is a corresponding multiplet of anomalies. For $\mathcal N=(1,0)$ theory, the `t~Hooft anomalies are characterized by the anomaly polynomial
\begin{equation}
    \mathcal I_8=\fft1{4!}[\alpha c_2(R)^2+\beta c_2(R)p_1(T)+\gamma p_1(T)^2+\delta p_2(T)],
\end{equation}
and the relation to the Weyl anomaly coefficients has recently been worked out \cite{Cordova:2015fha,Beccaria:2015ypa,Yankielowicz:2017xkf,Beccaria:2017dmw}
\begin{equation}
    a=-\fft1{72}(\alpha-\beta+\gamma+\ft38\delta),\qquad c-a=-\fft\delta{192},\qquad c'=\fft1{432}(\beta-2\gamma+\ft12\delta).
\end{equation}
Since $\alpha$ is the coefficient of the $[SU(2)_R]^4$ anomaly, it can be at most fifth power in $\Delta$, where the extra power comes from the dimension of the representation.  Similarly, $\beta$ can be at most cubic in $\Delta$, while $\gamma$ and $\delta$ can be at most linear in $\Delta$.  This in turn demonstrates that $a$ will be at most fifth power in $\Delta$, $c'$ will be at most cubic and $c-a$ will be at most linear.  Thus the functional Schr\"odinger method is indeed compatible with $\delta(c-a)$.  However, we also see this approach cannot be used to compute either $\delta a$ alone or $\delta c'$.  Thus, while it would be desirable to compute $\delta c'$ in these theories, we do not expect that it can be done using this approach.

While we have focused on short multiplets with spins $\le2$, it would be desirable to work more generally with higher-spin multiplets.  To do so, we would need knowledge of the $b_6$ coefficients for arbitrary spin fields.  This in turn depends on the form of the higher-spin Laplacian.  In general, this depends on the bulk dynamics of the higher-spin field and the further restriction to the boundary following from the procedure of \cite{Mansfield:1999kk,Mansfield:2000zw,Mansfield:2002pa,Mansfield:2003gs}.  For higher-spin bosons, it is natural to take a bulk Laplacian of the form $\Delta=-\Box-E$ with the endomorphism $E=\Sigma_{ab}R^{abcd}\Sigma_{cd}$, where $\Sigma_{ab}$ are $SU(4)$ generators in the appropriate bosonic higher-spin representation.  However, the situation is less clear for fermions.  The natural generalization would be to simply take $\Sigma_{ab}$ to be in a fermionic higher-spin representation.  However, this does not agree with the square of the Dirac operator for ordinary spin-1/2 fermions.  Nevertheless, it is possible that the use of a universal endomorphism term for bosons and fermions would be appropriate when tracing over supermultiplets.  Along these lines, we have computed the $b_6$ coefficient for general higher-spin representations in Appendix~\ref{sec:appB}.

Finally, part of our motivation for exploring the $\mathcal O(1)$ contributions to the holographic Weyl anomaly is to make a connection to the $\mathcal N=(1,0)$ superconformal index.  As in the AdS$_5$/CFT$_4$ case, the bulk one-loop corrections to the Weyl anomaly vanish for long representations, so it is natural to expect that these corrections can be obtained from the index.  More generally, we anticipate that the one-loop matching between $\delta(c-a)$ and the index can be extended to the full set of anomaly coefficients $a$ and $c_i$, even in theories without holographic duals.  This would then open up a new path towards characterizing superconformal field theories in six dimensions.

\acknowledgments

We wish to thank A.~Arabi Ardehali, F.~Larsen and P.~Szepietowski for stimulating discussions.  This work was supported in part by the US Department of Energy under Grant No.~DE-SC0007859.  The work of B.~McPeak was partially supported by a Rackham 2016 Spring/Summer Research Grant.

\appendix

\section{Heat kernel for spins up to two}
\label{app:b6fields}

The Seeley-DeWitt coefficients $b_n(\Delta)$ depend on the field and the form of the second order operator $\Delta$.  In four dimensions, the appropriate operators for irreducible fields up to spin two are listed in \cite{Christensen:1978md}. Here we write down the analogous operators in six dimensions and compute the contribution of each to the anomaly.

We start with the basis of curvature invariants \cite{parker1987,Bastianelli:2000hi}
\begin{align}
& A_1 = \Box^2 R, \quad 
A_2 = \left(\nabla_a R \right)^2, \quad A_3 = \left(\nabla_a R_{m n} \right)^2, \quad 
A_4 = \nabla_a R_{b m} \nabla^b R^{a m}, \quad 
A_5 = \left(\nabla_a R_{m n i j} \right)^2,
\nn\\
& A_6 =  R \Box R, \quad
A_7 = R_{a b}\Box R^{a b}, \quad
A_8 = R_{a b} \nabla_m \nabla^b R^{a m}, \quad
A_9 = R_{a b m n}\Box R^{a b m n}, \quad
A_{10} = R^3
\nn\\
& A_{11} = R R_{ab}^2, \quad
A_{12} = R R_{abmn}^2, \quad
A_{13} = R_a^{\ m} R_m^{\ \, i} R_i^{\ a}, \quad
A_{14} = R_{ab} R_{mn} R^{ambn}, \quad
\nn\\
& A_{15}= R_{ab} R^{amnl} R^b_{\ \,mnl}, \quad
A_{16} = R_{ab}{}^{cd}R_{cd}{}^{ef}R_{ef}{}^{ab},\quad 
A_{17} = R_{aibj}R^{manb}R^i{}_m{}^j{}_n.
\end{align}
The $b_6$ coefficient may be computed from the expression (\ref{eq:b6coef}), where the $V_a$'s are given by
\begin{align}
    & 
    V_1 = \nabla_k F_{ij} \nabla^k F^{ij}, \quad
    V_2 = \nabla_j F_{ij} \nabla^k F^{ik}, \quad
    V_3 =  F_{ij} \Box F^{ij}, \quad
    V_4 = F_{ij} F^{jk} F_{k}{}^i, \quad
    \nn\\
    &
    V_5 = R_{mnij} F^{mn} F^{ij}, \quad
    V_6 = R_{jk} F^{jn} F^k{}_n, \quad
    V_7 = R F_{ij} F^{ij}, \quad
    V_8 = \Box^2 E, \quad
    V_9 = E \Box E, \quad
    \nn\\
    &
    V_{10} = \nabla_k E \nabla^k E, \quad
    V_{11} = E^3, \quad
    V_{12} = E F_{ij}^2, \quad
    V_{13} = R \Box E, \quad
    V_{14} = R_{i j} \nabla^i \nabla^j E, \quad
    \nn\\
    &
    V_{15} = \nabla_k R \nabla^k E, \quad
    V_{16} = E^2 R, \quad
    V_{17} = E \Box R, \quad
    V_{18} = E R^2, \quad
    \nn\\
    &
    V_{19} = E R_{ij}^2, \quad
    V_{20} = E R_{ijkl}^2.
\end{align}
Here $\Delta=-\nabla^2-E$ and $F_{ij}$ is the curvature of the connection, $[\nabla_i,\nabla_i]=F_{ij}$.

\subsection{Conformally Coupled Scalar}

The conformally coupled scalar has $E = -\frac{1}{5} R$ and $F_{ij} = 0$, so the $V$-terms are:
\begin{center}
$\begin{array}{| c | c | c | c | c | c | c | c | c | c |}
    \hline
   V_1 & V_2 & V_3 & V_4 & V_5 & V_6 & V_7 & V_8 & V_9 & V_{10}\\
   \hline
   0 & 0 & 0 & 0 & 0 & 0 & 0 &  -\frac{1}{5} A_1 & \frac{1}{25} A_6 & \frac{1}{25} A_2 \\
   \hline
   \hline
   V_{11} & V_{12} & V_{13} & V_{14} & V_{15} & V_{16} & V_{17} & V_{18} & V_{19} & V_{20} \\
   \hline 
    -\frac{1}{125} A_{10} & 0 & -\frac{1}{5} A_6 & \frac{2}{5} (-A_8 + A_{13} - A_{14}) & -\frac{1}{5} A_2 & \frac{1}{25} A_{10} & -\frac{1}{5} A_6 & -\frac{1}{5} A_{10} & -\frac{1}{5} A_{11} & -\frac{1}{5} A_{12}  \\
   \hline
\end{array}$
\end{center}
The $b_6$ coefficient is
\begin{align}
b_6(\mathcal{O}) &= \frac{1}{(4 \pi)^3 7!} \bigg[ \frac{6}{5} A_1 + \frac{1}{5} A_2 - 2 A_3 -4 A_4 + 9 A_5 - 8  A_7 + \frac{8}{5} A_8 + 12 A_9 \nn\\
    &\quad-\frac{7}{225} A_{10} + \frac{14}{15} A_{11} - \frac{14}{15} A_{12} - \frac{32}{45} A_{13} -\frac{16}{15} A_{14} -\frac{16}{3} A_{15} + \frac{44}{9} A_{16} +\frac{80}{9} A_{17} \bigg].
\end{align}

\subsection{Weyl Fermion}
The appropriate second order operator for the Dirac fermion may be obtained as the square of the Dirac operator:
\begin{align}
    \mathcal{O} \psi = -\Box \psi + \frac{1}{4} R  \psi.
\end{align}
The endomorphism and curvature of the connection coincide with the result obtained in \cite{Bastianelli:2000hi}. 
\begin{align}
    E = -\frac{1}{4} R , \qquad F_{i j} = \frac{1}{4} R_{i j a b} \gamma^{a  b} .
\end{align}
Then the $V$-terms contributing to the anomaly are:
\begin{center}
$\begin{array}{| c | c | c | c | c | c | c | c | c | c |}
    \hline
   V_1 & V_2 & V_3 & V_4 & V_5 & V_6 & V_7 & V_8 & V_9 & V_{10}\\
   \hline
   -\frac{1}{2} A_5 & A_4 - A_3 & -\frac{1}{2} A_9 & \frac{1}{2} A_{17} & -\frac{1}{2} A_{16} & -\frac{1}{2} A_{15} & -\frac{1}{2} A_{12} & -A_1 & \frac{1}{4} A_6 & \frac{1}{4} A_2 \\
   \hline
    \hline
   V_{11} & V_{12} & V_{13} & V_{14} & V_{15} & V_{16} & V_{17} & V_{18} & V_{19} & V_{20} \\
   \hline 
   -\frac{1}{16} A_{10} & \frac{1}{8} A_{12} & -A_6 & -2(A_8 -A_{13} + A_{14}) & -A_2 & \frac{1}{4} A_{10} & -A_6 & -A_{10} & -A_{11} & -A_{12} \\
   \hline
\end{array}$
\end{center}
The $b_6$ coefficient is
\begin{align}
b_6(\mathcal{O}) &= \frac{4}{(4 \pi)^3 7!} \bigg[ -3 A_1 + \frac{5}{4} A_2 - 9A_3 +3 A_4 -5 A_5 + \frac{7}{2} A_6 - 8 A_7 -4 A_8 -9 A_9 \nn\\
    &\quad- \frac{35}{72} A_{10} +\frac{7}{3} A_{11} + \frac{49}{24} A_{12} + \frac{44}{9} A_{13} -\frac{20}{3} A_{14} +\frac{5}{3} A_{15} - \frac{101}{18} A_{16} - \frac{109}{9} A_{17} \bigg].
\label{eq:b6-100}
\end{align}

\subsection{Vector}
The $(0, 1, 0)$ vector representation of $SU(4)$ is a one form, so the correct Laplacian may be obtained by computing the Hodge-deRham operator $d \delta + \delta d$. We get 
\begin{align}
    \mathcal{O} A_{\mu} = -\Box A_{\mu} + R^{\nu}_{\ \, \mu} A_{\nu}.
\end{align}
The endomorphism and curvature of the connection here are: 
\begin{align}
    E^{a}{}_{b} = -R^{a}{}_{b}, \qquad (F_{i j})^{a}{}_{b} = R^{a}{}_{b i j},
\end{align}
so that
\begin{center}
$\begin{array}{| c | c | c | c | c | c | c | c | c | c |}
    \hline
   V_1 & V_2 & V_3 & V_4 & V_5 & V_6 & V_7 & V_8 & V_9 & V_{10}\\
   \hline
   - A_5 & 2(A_4 - A_3) & - A_9 & A_{17} & - A_{16} & - A_{15} & - A_{12} & -A_1 & A_7 &  A_3 \\
   \hline
    \hline
   V_{11} & V_{12} & V_{13} & V_{14} & V_{15} & V_{16} & V_{17} & V_{18} & V_{19} & V_{20} \\
   \hline 
   - A_{13} & A_{15} & -A_6 & -2(A_8 -A_{13} + A_{14}) & -A_2 & A_{11} & -A_6 & -A_{10} & -A_{11} & -A_{12} \\
   \hline
\end{array}$
\end{center}
and
\begin{align}
b_6(\mathcal{O}) &= \frac{1}{(4 \pi)^3 7!} \bigg[ 24 A_1 - 66 A_2 + 352 9A_3 + 32 A_4 -58 A_5 -140 A_6 + 792 A_7 - 32 A_8 - 96 A_9 \nn\\
    &\quad- \frac{140}{3} A_{10} + 420 A_{11} -70 A_{12} - \frac{2600}{3} A_{13} + 16 A_{14} + 444 A_{15} - \frac{164}{3} A_{16} - \frac{344}{3} A_{17} \bigg].
\end{align}

\subsection{Self-Dual Three-Form}

The field which transforms under the $(2, 0, 0)$ representation is the 10-component self-dual three-form. A three-index antisymmetric tensor has 20 components and the self-duality condition removes half of these. The operator acting on this field is
\begin{align}
    \mathcal{O} C_{\mu \nu \rho} &= -\Box C_{\mu \nu \rho} + R_{\mu}{}^{\lambda}  C_{\lambda \nu \rho} + R_{\nu}{}^{\lambda}  C_{\mu \lambda \rho} + R_{\rho}{}^{\lambda}  C_{\mu \nu \lambda} \nn\\
    &\quad- R_{\mu \nu}{}^{\lambda \sigma} C_{\lambda \sigma \rho} - R_{\nu \rho}{}^{\lambda \sigma} C_{\mu \lambda \sigma}- R_{\rho \mu}{}^{\lambda \sigma} C_{ \lambda \nu \sigma}.
\end{align}
This means that the endomorphism and connection curvature are given by 
\begin{align}
    E^{\ \ \ def}_{abc} = -3 R^{\ [d}_{[a}\delta^{e}_{b} \, \delta^{f]}_{c]} + 3 R_{[a b}{}^{[d e} \, \delta^{f]}_{c]} ,
    \qquad (F_{i j})_{abc}{}^{def} = -3 R_{i j [a}{}^{[d}\delta^{e}_{b} \delta^{f]}_{c]}.
\end{align}
Then we can compute the relevant terms:
\begin{center}
$\begin{array}{| c | c | c | c | c |}
    \hline
    V_1 & V_2 & V_3 & V_4 & V_5 \\ 
    \hline 
    -6 A_5 & 12(A_4 - A_3) & -6 A_9 & 6 A_{17} & -6 A_{16} \\ 
    \hline
    \hline
    V_6 & V_7 & V_8 & V_9 & V_{10}\\
    \hline
    -6 A_{15} & -6 A_{12} & -6 A_1 & 2 A_6 -2 A_7 + 2 A_9 & 2 A_2 -2 A_3 + 2 A_5\\
    \hline
    \hline
    V_{11} & V_{12} & V_{13} & V_{14} & V_{15}\\ 
    \hline 
    * & 2 A_{12} -2 A_{15} + 2 A_{16} & - 6 A_6 & -12 (A_8 -A_{13} + A_{14}) & - 6 A_2\\ 
    \hline
    \hline
    V_{16} & V_{17} & V_{18} & V_{19} & V_{20}\\
    \hline
    2 (A_{10} -A_{11} + A_{12}) & - 6 A_6 & - 6 A_{10} & - 6 A_{11} & - 6 A_{12}\\ 
    \hline
\end{array}$
\end{center}
where $V_{11}=- A_{10} +6 A_{11} -3 A_{12} -6A_{13} - 12A_{14} + 12 A_{15} - 2 A_{16} + 8 A_{17}$.  So the $b_6$ coefficient is given by
\begin{align}
b_6(\mathcal{O}) = \frac{1}{(4 \pi)^3 7!} \bigg[& -144 A_1 +172 A_2 -1216 A_3 + 256 A_4 + 348 A_5 + 392 A_6 - 1840 A_7 \nn\\
&- 192 A_8 + 912 A_9 
    + \frac{3080}{9} A_{10} +\frac{12824}{3} A_{11} - \frac{4004}{3} A_{12} - \frac{43472}{9} A_{13} \nn\\ &-\frac{30976}{3} A_{14} +\frac{28408}{3} A_{15} - \frac{11216}{9} A_{16} + \frac{53008}{9} A_{17} \bigg].
\end{align}
The self-duality condition reduces each of these terms by a factor of two, reproducing the $A_{16}$ and $A_{17}$ terms found in table~\ref{tbl:b6coef}.

\subsection{Gravitino}
The gravitino with the gauge condition $\gamma^{\mu} \psi_{\mu}=0$ corresponds to the $(1,1,0)$ representation. In this case the operator $\mathcal{O}$ is the square of the Rarita-Schwinger operator: 
\begin{align}
    \mathcal{O} \psi_{\mu} = -\Box \psi_{\mu} + \frac{1}{4} R  \psi_{\mu} - \frac{1}{2}\gamma^{\rho}\gamma^{\sigma} R_{\rho \sigma \mu \nu}  \psi^{\nu}.
\end{align}
The endomorphism and connection curvature are given by 
\begin{align}
    E_{b}{}^{a} = -\frac{1}{4} R \delta^{a}_{b} + \frac{1}{2} R_{c d b}{}^{a} \gamma^{c d},\qquad (F_{i j})^{\ \ a}_{b} = \frac{1}{4} R_{i j c d} \gamma^{c d} \delta^{a}_{b} +  R_{i j b}{}^{a} ,
\end{align}
so
\begin{center}
$\begin{array}{| c | c | c | c | c |}
    \hline
   V_1 & V_2 & V_3 & V_4 & V_5\\ 
   \hline 
   -7 A_{15} & -7 A_{12} & -6 A_1 & \frac{3}{2} A_6 + 2A_9 & \frac{3}{2} A_2 + 2A_5 \\ 
   \hline
   \hline
   V_6 & V_7 & V_8 & V_9 & V_{10}\\
   \hline
   -7 A_5 & 14(A_4 - A_3) & -7 A_9 & 7 A_{17} & -7 A_{16}\\
   \hline
   \hline
   V_{11} & V_{12} & V_{13} & V_{14} & V_{15} \\ 
   \hline 
   -\frac{3}{8} A_{10} - \frac{3}{2} A_{12} + 4A_{17} & \frac{7}{4} A_{12} + 2 A_{16} & - 6 A_6 & -12 (A_8 -A_{13} + A_{14}) & - 6 A_2\\ 
   \hline\hline
   V_{16} & V_{17} & V_{18} & V_{19} & V_{20} \\
   \hline 
   \frac{3}{2} A_{10} + 2A_{12} & - 6 A_6 & - 6 A_{10} & - 6 A_{11} & - 6 A_{12} \\
   \hline
\end{array}$
\end{center}
and
\begin{align}
b_6(\mathcal{O}) = \frac{1}{(4 \pi)^3 7!} \bigg[ & -60 A_1 + 25 A_2 - 404 A_3 + 284 A_4 + 292 A_5 + 70 A_6 - 160 A_7 \nn\\ 
    &- 80 A_8 +828 A_9 
    - \frac{175}{18} A_{10} +\frac{140}{3} A_{11} - \frac{1435}{6} A_{12} - \frac{880}{9} A_{13} \nn\\ & -\frac{400}{3} A_{14} +\frac{772}{3} A_{15} + \frac{3526}{9} A_{16} + \frac{22012}{9} A_{17} \bigg].
\label{eq:b6-110}
\end{align}

\subsection{Two-Form}
The adjoint representation $(1, 0, 1)$ corresponds to the two-form computed in \cite{Bastianelli:2000hi}. 
\begin{align}
    \mathcal{O} B_{\mu \nu} = -\Box B_{\mu \nu} + R^{\lambda}_{\mu}  B_{\lambda \nu} - R^{\lambda}_{\nu}  B_{\lambda \mu} - R^{\ \ \ \ \rho \sigma}_{\mu \nu} B_{\rho \sigma} .
\end{align}
This means that the endomorphism and connection curvature are given by 
\begin{align}
    E_{ab}{}^{cd} = -2 R^{[c}_{[a}\delta^{d]}_{b]} + R_{ab}{}^{cd} , \qquad (F_{i j})_{ab}{}^{cd} = 2R_{ij[a}{}^{[c} \delta^{d]}_{b]},
\end{align}
so
\begin{center}
$\begin{array}{| c | c | c | c | c |}
    \hline
    V_1 & V_2 & V_3 & V_4 & V_5 \\
    \hline
    -4 A_5 & 8(A_4 - A_3) & -4 A_9 & 4 A_{17} & -4 A_{16} \\
    \hline
    \hline
    V_6 & V_7 & V_8 & V_9 & V_{10}\\
    \hline
    -4 A_{15} & -4 A_{12} & -4 A_1 & A_6 + A_9 & A_2 + A_5 \\
    \hline
    \hline
    V_{11} & V_{12} & V_{13} & V_{14} & V_{15}\\
    \hline 
    -3 A_{11} + 4 A_{13} + 6 A_{14} -6 A_{15} + A_{16} & A_{12} + A_{16} & - 4 A_6 & -8 (A_8 -A_{13} + A_{14}) & - 4 A_2 \\
    \hline 
    \hline
    V_{16} & V_{17} & V_{18} & V_{19} & V_{20} \\
   \hline A_{10} + A_{12} & - 4 A_6 & - 4 A_{10} & - 4 A_{11} & - 4 A_{12} \\
   \hline
\end{array}$
\end{center}
and
\begin{align}
b_6(\mathcal{O}) &= \frac{1}{(4 \pi)^3 7!} \bigg[ -66 A_1 + 3 A_2 - 254 A_3 + 164 A_4 + 107 A_5 + 28 A_6 - 120 A_7 - 88 A_8 + 348 A_9 \nn\\
    &\kern-1em+ \frac{595}{3} A_{10} -2478 A_{11} + 518 A_{12} + \frac{10384}{3} A_{13} -4912 A_{14} - 4896 A_{15} + \frac{2992}{3} A_{16} - \frac{1616}{3} A_{17} \bigg].
\end{align}

\subsection{Graviton}

The symmetric spin-two field is the $(0, 2, 0)$ representation. The appropriate kinetic operator is the Lichnerowicz operator \cite{Lich}:
\begin{align}
    \mathcal{O} h_{\mu \nu} = -\Box h_{\mu \nu} + R_{\mu}{}^{\lambda} h_{\lambda \nu} + R_{\nu}{}^{\lambda} h_{\lambda \mu} - 2 R_{\mu \rho \nu \sigma} h^{\rho \sigma}.
\end{align}
The endomorphism and connection are given by
\begin{align}
    E_{\mu \nu}^{ \rho \sigma} = -2 R_{ \{ \mu}^{\ \{ \rho} \delta_{\nu \} }^{\ \sigma \} } + R_{\mu \ \ \nu}^{\ \ \rho \ \ \sigma} + R_{\mu \ \ \nu}^{\ \ \sigma \ \ \rho}, \qquad (F_{a b})_{\mu \nu}^{ \rho \sigma} = 2 R_{a b \{ \mu}^{\ \ \ \ \{ \rho} \delta_{\nu \}}^{\ \sigma \}}.
\end{align}
Then we can compute the relevant terms:
\begin{center}
$\begin{array}{| c | c | c | c | c |}
        \hline
   V_1 & V_2 & V_3 & V_4 & V_5 \\
   \hline
   -8 A_5 & 16(A_4 - A_3) & -8 A_9 & 8 A_{17} & -8 A_{16} \\
   \hline
   \hline
   V_6 & V_7 & V_8 & V_9 & V_{10}\\
   \hline
   -8 A_{15} & -8 A_{12} & -8 A_1 & A_6 + 12 A_7 + 3 A_9 & A_2 + 12 A_3 + 3 A_5 \\
   \hline
   \hline
   V_{11} & V_{12} & V_{13} & V_{14} & V_{15} \\
    \hline 
   * &  A_{12} + 12 A_{15} +3 A_{16}  & - 8 A_6  & -16 (A_8 -A_{13} + A_{14})  & - 8 A_2  \\
   \hline
   \hline
   V_{16} & V_{17} & V_{18} & V_{19} & V_{20} \\
   \hline
   A_{10} +12 A_{11} + 3A_{12}  & - 8 A_6  & - 8 A_{10}  & - 8 A_{11}  & - 8 A_{12} \\
   \hline
\end{array}$
\end{center}
where $V_{11}=-3A_{11} -16 A_{13} -6A_{14}  -18 A_{15} -A_{16} + 8A_{17}$. The $b_6$ coefficient is
\begin{align}
b_6(\mathcal{O}) = \frac{1}{(4 \pi)^3 7!} \bigg[ & -312 A_1 -584 A_2 - 4552 A_3 + 368 A_4 + 544 A_5 - 1064 A_6 + 9920 A_7 \nn\\ &- 416 A_8 + 1416  A_9 
    - \frac{560}{9} A_{10} +\frac{7952}{3} A_{11} + \frac{2968}{3} A_{12} - \frac{117056}{9} A_{13} \nn\\ &-\frac{16528}{3} A_{14} -\frac{29216}{3} A_{15} - \frac{1388}{9} A_{16} + \frac{49984}{9} A_{17} \bigg].
\end{align}

\section{Heat kernel for general spins}
\label{sec:appB}

We are interested in a general formula to compute the heat kernel coefficients for spins higher than two, analogous to the algorithm \cite{Christensen:1978md} in four dimensions. We consider fields transforming in an irreducible representation of the spacetime symmetry group that are acted on by a generalized second-order operator $\Delta = -\Box - E$. In four dimensions, the method of computing the heat kernels for general representations assumes that the endomorphism term $E$ for fields transforming as $(A,B)$ of $SO(4)\simeq SU(2)_L\times SU(2)_R$ is given by:
\begin{align}
    E = \Sigma_{ab}R^{abcd}\Sigma_{cd}\qquad\mbox{or} \qquad E = \frac{1}{A}\Sigma_{ab}R^{abcd}_{+}\Sigma_{cd},
\end{align}
for bosonic ($A + B =$ integer) or fermionic ($A + B =$ half-integer, $A>B$) representations, respectively. Here $R^{abcd}_{+} = \frac{1}{2}(R^{abcd} + R^{* \ abcd})$. This prescription is shown to be valid for fields up to spin two in four dimensions, and is conjectured to be the appropriate operator for general spins. In six dimensions, it appears that this prescription is reasonable for bosonic representations, but straightforward generalizations for fermions fail to reproduce the conventional endomorphism terms for the Weyl fermion and gravitino. So it remains unclear what endomorphism term is appropriate for general fermions. Below we use this method for bosonic representations to compute all the $V$ terms, which are built out of the endomorphism $E$ and the connection $F_{ij}$. 
\subsection{Tracing Over Generators}

Computing the heat kernel using this method requires computing the trace of a number of generators; the most we will need is six, as $E^3 \sim \Sigma^6$. We perform these traces using the algorithm presented in \cite{GroupTheory}, which requires expanding the trace into a sum of symmetric traces, and then writing each symmetric trace in a basis of orthogonal tensors and higher order Dynkin indices. For example, the trace of two generators of an irreducible representation is 
\begin{align}
    \text{Tr}[T_R^A T_R^B] = I_2(R) g^{AB}.
\end{align}
Here $R$ refers to the representation, and the capital Roman letters $A, B, \ldots=1,2,\ldots,15$ label the generators of $SU(4)$. Each $SU(4)$ index is interchangeable with a pair of antisymmetrized six-dimensional spacetime indices $\{\mu, \nu\}$.

If the number of generators is greater than two, we will first need to break the trace into a sum of symmetrized traces. For a trace of $n$ generators, this is accomplished by writing out each of the $n!$ terms in the symmetrized trace, and then using commutation relations to return each term to the original order, plus a number of traces of lower numbers of generators. For example, we may look at the trace of six generators. First consider the symmetrized trace
\begin{align}
    \text{STr} & [ T_{A}  T_{B} T_{C}  T_{D} T_{E}  T_{F}] \nn\\
    &= \frac{1}{6!} \left(    \text{Tr}[ T_{A}  T_{B} T_{C}  T_{D} T_{E}  T_{F}] +     \text{Tr}[ T_{B}  T_{A} T_{C}  T_{D} T_{E}  T_{F}] + 718 \text{ more terms} \right).
\end{align}
Using the fact that $T_B T_A = [T_B, T_A] + T_A T_B$ and the algebra, we may rewrite this  trace as
\begin{align}
        \text{STr} & [ T_{A}  T_{B} T_{C}  T_{D} T_{E}  T_{F}]
    = \frac{1}{6!} (    \text{Tr}[ T_{A}  T_{B} T_{C}  T_{D} T_{E}  T_{F}] 
    \nn\\
    & \quad +     \text{Tr}[ T_{A}  T_{B} T_{C}  T_{D} T_{E}  T_{F}] +     \text{Tr}[ f_{BAX} T^X T_{C}  T_{D} T_{E}  T_{F}] + 718 \text{ more terms}).
\end{align}
This gives two factors of the non-symmetrized trace plus a term which has a trace over only five generators. Each of the other 718 terms may be dealt with in the same way: commute the generators to put them in the order $(ABCDEF)$ and keep track of all of the traces over five generators which are picked up along the way. This adds $5\cdot5!$ terms with five generators. Using this and rearranging the trace and symmetric trace, we get the schematic relation
\begin{equation}
    \text{Tr} [ T_{A}  T_{B} T_{C}  T_{D} T_{E}  T_{F}]
    = \text{STr}[ T_{A}  T_{B} T_{C}  T_{D} T_{E}  T_{F}] -  \frac{1}{6!}\cdot600 \, \text{Tr}[TTTTT].
\end{equation}
Each of these five-generator traces may be treated the same way-- they each yield a symmetric trace with five generators plus $4\cdot4!$ terms with a trace over four generators. Schematically, the trace may be expanded as
\begin{align}
    \text{Tr} [ T_{A}  T_{B} T_{C}  T_{D} T_{E}  T_{F}] &\nn\\
    &\kern-4em= \text{STr}[ T_{A}  T_{B} T_{C}  T_{D} T_{E}  T_{F}] -  \frac{1}{6!} \left( 600 \Big( \text{STr}[TTTTT] - \frac{1}{5!}\cdot 96 \, \text{Tr}[TTTT] \Big) \right),
\end{align}
and so on, until the result is a sum of symmetric traces of 2, 3, 4, 5, and 6 generators. Clearly this computation is not tractable by hand. Using the XACT package for Mathematica, we calculated all the necessary terms. The symmetric traces over an odd number of generators cancel each other out (which appears to be a sort of generalization of Furry's theorem).  The result of this procedure includes a symmetric trace over six generators and a large number of symmetric traces over four generators and two generators.

\subsection{Orthogonal Tensors}

The symmetrized traces may be expanded in a set of orthogonal symmetric tensors. The two needed for this calculation are
\begin{align}
    \text{STr}[T^A T^B T^C T^D] =& I_4 (R) d_{\bot}^{ABCD} + I_{2,2}(R) (\delta^{AB} \delta^{CD} + \delta^{AC} \delta^{BD} +\delta^{AD} \delta^{BC})/3,
\end{align}
and
\begin{align}
    \text{STr} [T^A T^B T^C T^D T^E T^F] &= I_6 (R) d_{\bot}^{ABCDEF} + I_{4,2}(R) (d_{\bot}^{ABCD} \delta^{EF} + d_{\bot}^{ABCE} \delta^{DF} + \cdots)/15 \nn\\
    &\kern-6em+ I_{3,3}(R) (d_{\bot}^{ABC}d_{\bot}^{DEF} + d_{\bot}^{ABD}d_{\bot}^{CEF} +\cdots)/10 + I_{2,2,2}(R) (\delta^{AB} \delta^{CD}\delta^{EF} + \cdots)/15.
\end{align}
Note that $I_6 = 0$ for all representations of $SU(4)$. The tensors $d_{\bot}^{ABCD}$ and $d_{\bot}^{ABC}$ are fixed by the condition of orthogonality; $d_{\bot}^{ABC}$ is the six-dimensional epsilon tensor (recalling that $A = \{\mu_1 \nu_1\}$, etc.) The fourth order $d_{\bot}^{ABCD}$ may be expressed in terms of the six-dimensional metric --- its terms include $g^{\mu_1 \nu_4 } g^{\mu_2 \nu_3} g^{\mu_3 \nu_2} g^{\mu_4 \nu_1}$ and the other 47 ways of arranging the indices. 
The indices
$I_{4,2}$, $I_{3,3}$, and $I_{2,2,2}$ are not unique; imposing orthogonality and other group-theoretic relations yields the system of equations (158)--(160) in \cite{GroupTheory}. Solving these allows $I_{4,2}$, $I_{3,3}$, and $I_{2,2,2}$ to be expressed in terms of the Dynkin indices $I_4$, $I_3$, and $I_2$.

\subsection{Dynkin Indices}
A representation $R$ with Dynkin labels $(a,b,c)$ has dimension
\begin{align}
    &\text{Dim}_R (a, b, c) = \frac{1}{12} (a+1) (b+1) (c+1) (a+b+2) (b+c+2) (a+b+c+3) .
\end{align}
The Weyl character formula may be used to show that
\begin{align}
    I_2(a,b,c) = \frac{\text{Dim}_R}{60} \left(3 a^2+2 a (2 b+c+6)+4 b^2+4 b (c+4)+3 c
   (c+4)\right).
\end{align}
The third and fourth order generalization to this index were computed in \cite{Okubo:1981td}, which finds
\begin{align}
   I_3(a,b,c) &= \frac{\text{Dim}_R}{120} (a-c) (a+c+2) (a+2 b+c+4)
   \nn\\
   I_4(a,b,c) &= \frac{\text{Dim}_R}{3360}
   \big( 3 a^4+8 a^3 b+4 a^3 c+24 a^3+2 a^2 b^2+2 a^2 b c+30 a^2 b
   \nn\\
   &\quad -4 a^2 c^2+6 a^2 c +54 a^2 -12 a b^3-18 a b^2 c-50 a b^2+2 a b c^2-28 a b c
   \nn\\
   &\quad -34 a b+4 a c^3+6 a c^2-2 a c+24 a -6 b^4-12 b^3 c-48 b^3+2 b^2 c^2
   \nn \\
   &\quad-50 b^2 c-122 b^2+8 b c^3+30 b c^2-34 b c-104 b+3 c^4+24 c^3+54 c^2+24  c\big).
\end{align}


\subsection{Results}
As each of the $V_a$ coefficients may be reduced to a trace of generators variously contracted with the Riemann tensor, this method will allow each of them to be computed. The entire list of coefficients is presented here:
\begin{align}
    V_1 &= - \frac{A_5}{2} I_2, \qquad
    V_2 = (A_4- A_3) I_2, \qquad
    V_3 = - \frac{A_9}{2} I_2, \qquad
    V_4 =  \frac{A_{17}}{2}  I_2\nn\\
    V_5 &= - \frac{A_{16}}{2}  I_2,\qquad
    V_6 = - \frac{A_{15}}{2}  I_2, \qquad
    V_7 = - \frac{A_{12}}{2}  I_2, \qquad
    V_8 = - \frac{A_1}{2}  I_2,\nn\\
    V_9 &= \left(-\frac{A_6}{51} + \frac{A_7}{6} - \frac{25 A_9}{204} \right) I_2 + \left( \frac{15 A_6}{68} + \frac{15 A_9}{34} \right) \frac{I_2^2}{\text{Dim}_R} + \left( \frac{11 A_6}{51} - \frac{4 A_7}{3} + \frac{5 A_9}{51}\right) I_4,\nn \\
    V_{10} &= \left(-\frac{A_2}{51} + \frac{A_3}{6} - \frac{25 A_5}{204} \right) I_2 + \left( \frac{15 A_2}{68} + \frac{15 A_5}{34} \right) \frac{I_2^2}{\text{Dim}_R} + \left( \frac{11 A_2}{51} - \frac{4 A_3}{3} + \frac{5 A_5}{51}\right) I_4, \nn\\
    V_{11} &= 
    \left( \frac{A_{10}}{612} - \frac{11 A_{11}}{357}-\frac{3A_{12}}{238} - \frac{55 A_{13}}{2142} + \frac{151 A_{14}}{714} + \frac{3 A_{15}}{34} - \frac{383 A_{16}}{4284} - \frac{338 A_{17}}{1071}
    \right) I_2\nn\\
    & \quad +\left( \frac{5 A_{10}}{136}-\frac{375
   A_{11}}{952}+\frac{1095
   A_{12}}{3808}+\frac{115
   A_{13}}{476}+\frac{345
   A_{14}}{952}-\frac{165
   A_{15}}{136}+\frac{325
   A_{16}}{476}+\frac{725
   A_{17}}{952}
    \right) \frac{I_2^2}{\text{Dim}_R}\nn\\
     & \quad+\left( \frac{10 A_{10}}{153}-\frac{41
   A_{11}}{51}+\frac{6
   A_{12}}{17}+\frac{280
   A_{13}}{153}+\frac{38
   A_{14}}{51}-\frac{42
   A_{15}}{17}+\frac{43
   A_{16}}{153}-\frac{8 A_{17}}{153}
    \right) I_4\nn\\
    & \quad +\left( -\frac{5 A_{10}}{68}-\frac{165
   A_{11}}{952}-\frac{1845
   A_{12}}{3808}+\frac{115
   A_{13}}{476}+\frac{345
   A_{14}}{952}-\frac{45
   A_{15}}{136}-\frac{305
   A_{16}}{476}-\frac{115
   A_{17}}{952}
    \right)\frac{I_2^3}{\text{Dim}_R^2}\nn
    \\
    & \quad +\left( -\frac{7 A_{10}}{24}+\frac{209
   A_{11}}{56}-\frac{183
   A_{12}}{224}-\frac{437
   A_{13}}{84}-\frac{437
   A_{14}}{56}+\frac{57
   A_{15}}{8}-\frac{101
   A_{16}}{84}+\frac{437 A_{17}}{168}
    \right)\frac{I_3^2}{\text{Dim}_R}
    \nn\\
    & \quad +\left( -\frac{13 A_{10}}{102}+\frac{4
   A_{11}}{17}-\frac{12
   A_{12}}{17}+\frac{76
   A_{13}}{51}+\frac{38
   A_{14}}{17}+\frac{54
   A_{15}}{17}-\frac{2
   A_{16}}{51}-\frac{38 A_{17}}{51}
    \right)\frac{I_2 I_4}{\text{Dim}_R},\nn\\
    V_{12} &= \left(-\frac{A_{12}}{51} + \frac{A_{15}}{6} - \frac{25 A_{16}}{204} \right) I_2 + \left( \frac{15 A_{12}}{68} + \frac{15 A_{16}}{34} \right) \frac{I_2^2}{\text{Dim}_R} + \left( \frac{11 A_{12}}{51} - \frac{4 A_{15}}{3} + \frac{5 A_{16}}{51}\right) I_4,\nn\\
    V_{13} &= -\frac{A_6}{2} I_2, \qquad
    V_{14} = -\left( A_8 -A_{13} + A_{14} \right) I_2, \qquad
    V_{15} = -\frac{A_2}{2} I_2,\nn\\
    V_{16} &= \left(-\frac{A_{10}}{51} + \frac{A_{11}}{6} - \frac{25 A_{12}}{204} \right) I_2 + \left( \frac{15 A_{10}}{68} + \frac{15 A_{12}}{34} \right) \frac{I_2^2}{\text{Dim}_R} + \left( \frac{11 A_{10}}{51} - \frac{4 A_{11}}{3} + \frac{5 A_{12}}{51}\right) I_4,\nn\\
    V_{17} &= -\frac{A_6}{2} I_2, \qquad
    V_{18} = -\frac{A_{10}}{2} I_2, \qquad
    V_{19} = -\frac{A_{11}}{2} I_2, \qquad
    V_{20} = -\frac{A_{12}}{2} I_2.
\end{align}
Since these expressions pertain to an endomorphism of the form $E=\Sigma_{ab}R^{abcd}\Sigma_{cd}$, where $\Sigma_{ab}$ are $SU(4)$ generators in an arbitrary representation specified by Dynkin labels $(a,b,c)$, we refer to this as the ``group theory method'' for determining the heat kernel coefficients.

Now that the $V_a$'s are known, we may compute the $b_6$ coefficient using the group theory method. We present the coefficient for a representation $R$ on Ricci-flat backgrounds:
\begin{align}
    b_6(R)\Big|_{R_{ab} = 0} &= \frac{1}{(4 \pi)^3 7!} \Bigg[ A_5 \left(\frac{3150
   I_2^2}{17
   \text{Dim}_R}+9
   \text{Dim}_R-\frac{1827
   I_2}{17}+\frac{700
   I_4}{17}\right)
   \nn\\
   &\kern5em + A_9 \left(\frac{6300
   I_2^2}{17
   \text{Dim}_R}+12
   \text{Dim}_R-\frac{3178
   I_2}{17}+\frac{1400
   I_4}{17}\right) \nonumber
    \\
    &\kern-4.5em + A_{16} \left(-\frac{9150
   I_2^3}{17
   \text{Dim}_R^2}+\frac{12900
   I_2^2}{17
   \text{Dim}_R}-\frac{560
   I_2 I_4}{17
   \text{Dim}_R}-\frac{1010
   I_3^2}{\text{Dim}_R}+\frac{44
   \text{Dim}_R}{9}-\frac{8597
   I_2}{51}+\frac{14140
   I_4}{51}\right) \nonumber
   \\
   &\kern-4.5em+A_{17}\left(-\frac{1725
   I_2^3}{17
   \text{Dim}_R^2}+\frac{10875
   I_2^2}{17
   \text{Dim}_R}-\frac{10640
   I_2 I_4}{17
   \text{Dim}_R}+\frac{2185
   I_3^2}{\text{Dim}_R}+\frac{80
   \text{Dim}_R}{9}-\frac{17804
   I_2}{51}-\frac{2240
   I_4}{51}\right) \Bigg].
\label{eq:b6gen}
\end{align}
In general, the full $b_6$ coefficients obtained by the group theory method do not match the expressions (\ref{eq:b6-100}) and (\ref{eq:b6-110}), for the fermion and gravitino, respectively, as the group theory method does not correspond to the square of the Dirac operator when acting on fermions. This indicates that some modification may be necessary for fermionic representations, as was already noted in the four-dimensional case \cite{Christensen:1978md}.
Curiously, however, this mismatch disappears when restricted to Ricci-flat backgrounds.  This suggests that (\ref{eq:b6gen}) may potentially be valid for fermions as well as bosons.  If this were true, we could then derive a general expression for $\delta(c-a)$ for arbitrary higher spin supermultiplets.

Finally, the expression $\delta\mathcal A$ in (\ref{eq:deltaA}) vanishes in arbitrary backgrounds for long multiplets using the group theory method for the heat kernel.  This is in contrast to the conventional method where the fermions are treated by squaring the Dirac operator.  There, $\delta\mathcal A$ for long multiplets only vanished on Ricci-flat backgrounds, but was otherwise non-vanishing on more general backgrounds.


\bibliography{cite.bib}

\begin{thebibliography}{35}%
\makeatletter
\providecommand \@ifxundefined [1]{%
 \@ifx{#1\undefined}
}%
\providecommand \@ifnum [1]{%
 \ifnum #1\expandafter \@firstoftwo
 \else \expandafter \@secondoftwo
 \fi
}%
\providecommand \@ifx [1]{%
 \ifx #1\expandafter \@firstoftwo
 \else \expandafter \@secondoftwo
 \fi
}%
\providecommand \natexlab [1]{#1}%
\providecommand \enquote  [1]{``#1''}%
\providecommand \bibnamefont  [1]{#1}%
\providecommand \bibfnamefont [1]{#1}%
\providecommand \citenamefont [1]{#1}%
\providecommand \href@noop [0]{\@secondoftwo}%
\providecommand \href [0]{\begingroup \@sanitize@url \@href}%
\providecommand \@href[1]{\@@startlink{#1}\@@href}%
\providecommand \@@href[1]{\endgroup#1\@@endlink}%
\providecommand \@sanitize@url [0]{\catcode `\\12\catcode `\$12\catcode
  `\&12\catcode `\#12\catcode `\^12\catcode `\_12\catcode `\%12\relax}%
\providecommand \@@startlink[1]{}%
\providecommand \@@endlink[0]{}%
\providecommand \url  [0]{\begingroup\@sanitize@url \@url }%
\providecommand \@url [1]{\endgroup\@href {#1}{\urlprefix }}%
\providecommand \urlprefix  [0]{URL }%
\providecommand \Eprint [0]{\href }%
\providecommand \doibase [0]{http://dx.doi.org/}%
\providecommand \selectlanguage [0]{\@gobble}%
\providecommand \bibinfo  [0]{\@secondoftwo}%
\providecommand \bibfield  [0]{\@secondoftwo}%
\providecommand \translation [1]{[#1]}%
\providecommand \BibitemOpen [0]{}%
\providecommand \bibitemStop [0]{}%
\providecommand \bibitemNoStop [0]{.\EOS\space}%
\providecommand \EOS [0]{\spacefactor3000\relax}%
\providecommand \BibitemShut  [1]{\csname bibitem#1\endcsname}%
\let\auto@bib@innerbib\@empty
\bibitem [{\citenamefont {Cardy}(1986)}]{Cardy:1986ie}%
  \BibitemOpen
  \bibfield  {author} {\bibinfo {author} {\bibfnamefont {John~L.}\ \bibnamefont
  {Cardy}},\ }\bibfield  {title} {\enquote {\bibinfo {title} {{Operator Content
  of Two-Dimensional Conformally Invariant Theories}},}\ }\href {\doibase
  10.1016/0550-3213(86)90552-3} {\bibfield  {journal} {\bibinfo  {journal}
  {Nucl. Phys.}\ }\textbf {\bibinfo {volume} {B270}},\ \bibinfo {pages}
  {186--204} (\bibinfo {year} {1986})}\BibitemShut {NoStop}%
\bibitem [{\citenamefont {Zamolodchikov}(1986)}]{Zamolodchikov:1986gt}%
  \BibitemOpen
  \bibfield  {author} {\bibinfo {author} {\bibfnamefont {A.~B.}\ \bibnamefont
  {Zamolodchikov}},\ }\bibfield  {title} {\enquote {\bibinfo {title}
  {{Irreversibility of the Flux of the Renormalization Group in a 2D Field
  Theory}},}\ }\href@noop {} {\bibfield  {journal} {\bibinfo  {journal} {JETP
  Lett.}\ }\textbf {\bibinfo {volume} {43}},\ \bibinfo {pages} {730--732}
  (\bibinfo {year} {1986})},\ \bibinfo {note} {[Pisma Zh. Eksp. Teor.
  Fiz.43,565(1986)]}\BibitemShut {NoStop}%
\bibitem [{\citenamefont {Henningson}\ and\ \citenamefont
  {Skenderis}(1998)}]{Henningson:1998gx}%
  \BibitemOpen
  \bibfield  {author} {\bibinfo {author} {\bibfnamefont {M.}~\bibnamefont
  {Henningson}}\ and\ \bibinfo {author} {\bibfnamefont {K.}~\bibnamefont
  {Skenderis}},\ }\bibfield  {title} {\enquote {\bibinfo {title} {{The
  Holographic Weyl anomaly}},}\ }\href {\doibase 10.1088/1126-6708/1998/07/023}
  {\bibfield  {journal} {\bibinfo  {journal} {JHEP}\ }\textbf {\bibinfo
  {volume} {07}},\ \bibinfo {pages} {023} (\bibinfo {year} {1998})},\ \Eprint
  {http://arxiv.org/abs/hep-th/9806087} {arXiv:hep-th/9806087 [hep-th]}
  \BibitemShut {NoStop}%
\bibitem [{\citenamefont {Bilal}\ and\ \citenamefont
  {Chu}(1999)}]{Bilal:1999ph}%
  \BibitemOpen
  \bibfield  {author} {\bibinfo {author} {\bibfnamefont {Adel}\ \bibnamefont
  {Bilal}}\ and\ \bibinfo {author} {\bibfnamefont {Chong-Sun}\ \bibnamefont
  {Chu}},\ }\bibfield  {title} {\enquote {\bibinfo {title} {{A Note on the
  chiral anomaly in the AdS / CFT correspondence and $1 / N^2$ correction}},}\
  }\href {\doibase 10.1016/S0550-3213(99)00553-2} {\bibfield  {journal}
  {\bibinfo  {journal} {Nucl. Phys.}\ }\textbf {\bibinfo {volume} {B562}},\
  \bibinfo {pages} {181--190} (\bibinfo {year} {1999})},\ \Eprint
  {http://arxiv.org/abs/hep-th/9907106} {arXiv:hep-th/9907106 [hep-th]}
  \BibitemShut {NoStop}%
\bibitem [{\citenamefont {Mansfield}\ and\ \citenamefont
  {Nolland}(1999)}]{Mansfield:1999kk}%
  \BibitemOpen
  \bibfield  {author} {\bibinfo {author} {\bibfnamefont {Paul}\ \bibnamefont
  {Mansfield}}\ and\ \bibinfo {author} {\bibfnamefont {David}\ \bibnamefont
  {Nolland}},\ }\bibfield  {title} {\enquote {\bibinfo {title} {{One loop
  conformal anomalies from AdS / CFT in the Schrodinger representation}},}\
  }\href {\doibase 10.1088/1126-6708/1999/07/028} {\bibfield  {journal}
  {\bibinfo  {journal} {JHEP}\ }\textbf {\bibinfo {volume} {07}},\ \bibinfo
  {pages} {028} (\bibinfo {year} {1999})},\ \Eprint
  {http://arxiv.org/abs/hep-th/9906054} {arXiv:hep-th/9906054 [hep-th]}
  \BibitemShut {NoStop}%
\bibitem [{\citenamefont {Mansfield}\ and\ \citenamefont
  {Nolland}(2000)}]{Mansfield:2000zw}%
  \BibitemOpen
  \bibfield  {author} {\bibinfo {author} {\bibfnamefont {Paul}\ \bibnamefont
  {Mansfield}}\ and\ \bibinfo {author} {\bibfnamefont {David}\ \bibnamefont
  {Nolland}},\ }\bibfield  {title} {\enquote {\bibinfo {title} {{Order $1 /
  N^2$ test of the Maldacena conjecture: Cancellation of the one loop Weyl
  anomaly}},}\ }\href {\doibase 10.1016/S0370-2693(00)01247-8} {\bibfield
  {journal} {\bibinfo  {journal} {Phys. Lett.}\ }\textbf {\bibinfo {volume}
  {B495}},\ \bibinfo {pages} {435--439} (\bibinfo {year} {2000})},\ \Eprint
  {http://arxiv.org/abs/hep-th/0005224} {arXiv:hep-th/0005224 [hep-th]}
  \BibitemShut {NoStop}%
\bibitem [{\citenamefont {Mansfield}\ \emph
  {et~al.}(2003{\natexlab{a}})\citenamefont {Mansfield}, \citenamefont
  {Nolland},\ and\ \citenamefont {Ueno}}]{Mansfield:2002pa}%
  \BibitemOpen
  \bibfield  {author} {\bibinfo {author} {\bibfnamefont {Paul}\ \bibnamefont
  {Mansfield}}, \bibinfo {author} {\bibfnamefont {David}\ \bibnamefont
  {Nolland}}, \ and\ \bibinfo {author} {\bibfnamefont {Tatsuya}\ \bibnamefont
  {Ueno}},\ }\bibfield  {title} {\enquote {\bibinfo {title} {{Order $1 / N^2$
  test of the Maldacena conjecture. 2. The Full bulk one loop contribution to
  the boundary Weyl anomaly}},}\ }\href {\doibase
  10.1016/S0370-2693(03)00750-0} {\bibfield  {journal} {\bibinfo  {journal}
  {Phys. Lett.}\ }\textbf {\bibinfo {volume} {B565}},\ \bibinfo {pages}
  {207--210} (\bibinfo {year} {2003}{\natexlab{a}})},\ \Eprint
  {http://arxiv.org/abs/hep-th/0208135} {arXiv:hep-th/0208135 [hep-th]}
  \BibitemShut {NoStop}%
\bibitem [{\citenamefont {Mansfield}\ \emph {et~al.}(2004)\citenamefont
  {Mansfield}, \citenamefont {Nolland},\ and\ \citenamefont
  {Ueno}}]{Mansfield:2003gs}%
  \BibitemOpen
  \bibfield  {author} {\bibinfo {author} {\bibfnamefont {Paul}\ \bibnamefont
  {Mansfield}}, \bibinfo {author} {\bibfnamefont {David}\ \bibnamefont
  {Nolland}}, \ and\ \bibinfo {author} {\bibfnamefont {Tatsuya}\ \bibnamefont
  {Ueno}},\ }\bibfield  {title} {\enquote {\bibinfo {title} {{The Boundary Weyl
  anomaly in the $\mathcal N=4$ SYM / type IIB supergravity correspondence}},}\
  }\href {\doibase 10.1088/1126-6708/2004/01/013} {\bibfield  {journal}
  {\bibinfo  {journal} {JHEP}\ }\textbf {\bibinfo {volume} {01}},\ \bibinfo
  {pages} {013} (\bibinfo {year} {2004})},\ \Eprint
  {http://arxiv.org/abs/hep-th/0311021} {arXiv:hep-th/0311021 [hep-th]}
  \BibitemShut {NoStop}%
\bibitem [{\citenamefont {Arabi~Ardehali}\ \emph {et~al.}(2013)\citenamefont
  {Arabi~Ardehali}, \citenamefont {Liu},\ and\ \citenamefont
  {Szepietowski}}]{Ardehali:2013gra}%
  \BibitemOpen
  \bibfield  {author} {\bibinfo {author} {\bibfnamefont {Arash}\ \bibnamefont
  {Arabi~Ardehali}}, \bibinfo {author} {\bibfnamefont {James~T.}\ \bibnamefont
  {Liu}}, \ and\ \bibinfo {author} {\bibfnamefont {Phillip}\ \bibnamefont
  {Szepietowski}},\ }\bibfield  {title} {\enquote {\bibinfo {title} {{The
  spectrum of IIB supergravity on AdS$_5 \times S^5/Z_3$ and a $1/N^2$ test of
  AdS/CFT}},}\ }\href {\doibase 10.1007/JHEP06(2013)024} {\bibfield  {journal}
  {\bibinfo  {journal} {JHEP}\ }\textbf {\bibinfo {volume} {06}},\ \bibinfo
  {pages} {024} (\bibinfo {year} {2013})},\ \Eprint
  {http://arxiv.org/abs/1304.1540} {arXiv:1304.1540 [hep-th]} \BibitemShut
  {NoStop}%
\bibitem [{\citenamefont {Ardehali}\ \emph {et~al.}(2014)\citenamefont
  {Ardehali}, \citenamefont {Liu},\ and\ \citenamefont
  {Szepietowski}}]{Ardehali:2013xya}%
  \BibitemOpen
  \bibfield  {author} {\bibinfo {author} {\bibfnamefont {Arash~Arabi}\
  \bibnamefont {Ardehali}}, \bibinfo {author} {\bibfnamefont {James~T.}\
  \bibnamefont {Liu}}, \ and\ \bibinfo {author} {\bibfnamefont {Phillip}\
  \bibnamefont {Szepietowski}},\ }\bibfield  {title} {\enquote {\bibinfo
  {title} {{$1/N^2$ corrections to the holographic Weyl anomaly}},}\ }\href
  {\doibase 10.1007/JHEP01(2014)002} {\bibfield  {journal} {\bibinfo  {journal}
  {JHEP}\ }\textbf {\bibinfo {volume} {1401}},\ \bibinfo {pages} {002}
  (\bibinfo {year} {2014})},\ \Eprint {http://arxiv.org/abs/1310.2611}
  {arXiv:1310.2611 [hep-th]} \BibitemShut {NoStop}%
\bibitem [{\citenamefont {Arabi~Ardehali}\ \emph
  {et~al.}(2014{\natexlab{a}})\citenamefont {Arabi~Ardehali}, \citenamefont
  {Liu},\ and\ \citenamefont {Szepietowski}}]{Ardehali:2013xla}%
  \BibitemOpen
  \bibfield  {author} {\bibinfo {author} {\bibfnamefont {Arash}\ \bibnamefont
  {Arabi~Ardehali}}, \bibinfo {author} {\bibfnamefont {James~T.}\ \bibnamefont
  {Liu}}, \ and\ \bibinfo {author} {\bibfnamefont {Phillip}\ \bibnamefont
  {Szepietowski}},\ }\bibfield  {title} {\enquote {\bibinfo {title} {{The
  shortened KK spectrum of IIB supergravity on $Y^{p,q}$}},}\ }\href {\doibase
  10.1007/JHEP02(2014)064} {\bibfield  {journal} {\bibinfo  {journal} {JHEP}\
  }\textbf {\bibinfo {volume} {02}},\ \bibinfo {pages} {064} (\bibinfo {year}
  {2014}{\natexlab{a}})},\ \Eprint {http://arxiv.org/abs/1311.4550}
  {arXiv:1311.4550 [hep-th]} \BibitemShut {NoStop}%
\bibitem [{\citenamefont {Beccaria}\ and\ \citenamefont
  {Tseytlin}(2014)}]{Beccaria:2014xda}%
  \BibitemOpen
  \bibfield  {author} {\bibinfo {author} {\bibfnamefont {Matteo}\ \bibnamefont
  {Beccaria}}\ and\ \bibinfo {author} {\bibfnamefont {Arkady~A.}\ \bibnamefont
  {Tseytlin}},\ }\bibfield  {title} {\enquote {\bibinfo {title} {{Higher spins
  in AdS$_{5}$ at one loop: vacuum energy, boundary conformal anomalies and
  AdS/CFT}},}\ }\href {\doibase 10.1007/JHEP11(2014)114} {\bibfield  {journal}
  {\bibinfo  {journal} {JHEP}\ }\textbf {\bibinfo {volume} {11}},\ \bibinfo
  {pages} {114} (\bibinfo {year} {2014})},\ \Eprint
  {http://arxiv.org/abs/1410.3273} {arXiv:1410.3273 [hep-th]} \BibitemShut
  {NoStop}%
\bibitem [{\citenamefont {Arabi~Ardehali}\ \emph
  {et~al.}(2014{\natexlab{b}})\citenamefont {Arabi~Ardehali}, \citenamefont
  {Liu},\ and\ \citenamefont {Szepietowski}}]{Ardehali:2014zba}%
  \BibitemOpen
  \bibfield  {author} {\bibinfo {author} {\bibfnamefont {Arash}\ \bibnamefont
  {Arabi~Ardehali}}, \bibinfo {author} {\bibfnamefont {James~T.}\ \bibnamefont
  {Liu}}, \ and\ \bibinfo {author} {\bibfnamefont {Phillip}\ \bibnamefont
  {Szepietowski}},\ }\bibfield  {title} {\enquote {\bibinfo {title} {{$c - a$
  from the $ \mathcal{N}=1 $ superconformal index}},}\ }\href {\doibase
  10.1007/JHEP12(2014)145} {\bibfield  {journal} {\bibinfo  {journal} {JHEP}\
  }\textbf {\bibinfo {volume} {12}},\ \bibinfo {pages} {145} (\bibinfo {year}
  {2014}{\natexlab{b}})},\ \Eprint {http://arxiv.org/abs/1407.6024}
  {arXiv:1407.6024 [hep-th]} \BibitemShut {NoStop}%
\bibitem [{\citenamefont {Arabi~Ardehali}\ \emph {et~al.}(2015)\citenamefont
  {Arabi~Ardehali}, \citenamefont {Liu},\ and\ \citenamefont
  {Szepietowski}}]{Ardehali:2014esa}%
  \BibitemOpen
  \bibfield  {author} {\bibinfo {author} {\bibfnamefont {Arash}\ \bibnamefont
  {Arabi~Ardehali}}, \bibinfo {author} {\bibfnamefont {James~T.}\ \bibnamefont
  {Liu}}, \ and\ \bibinfo {author} {\bibfnamefont {Phillip}\ \bibnamefont
  {Szepietowski}},\ }\bibfield  {title} {\enquote {\bibinfo {title} {{Central
  charges from the $\mathcal{N} =1$ superconformal index}},}\ }\href {\doibase
  10.1103/PhysRevLett.114.091603} {\bibfield  {journal} {\bibinfo  {journal}
  {Phys. Rev. Lett.}\ }\textbf {\bibinfo {volume} {114}},\ \bibinfo {pages}
  {091603} (\bibinfo {year} {2015})},\ \Eprint {http://arxiv.org/abs/1411.5028}
  {arXiv:1411.5028 [hep-th]} \BibitemShut {NoStop}%
\bibitem [{\citenamefont {Di~Pietro}\ and\ \citenamefont
  {Komargodski}(2014)}]{DiPietro:2014bca}%
  \BibitemOpen
  \bibfield  {author} {\bibinfo {author} {\bibfnamefont {Lorenzo}\ \bibnamefont
  {Di~Pietro}}\ and\ \bibinfo {author} {\bibfnamefont {Zohar}\ \bibnamefont
  {Komargodski}},\ }\bibfield  {title} {\enquote {\bibinfo {title} {{Cardy
  formulae for SUSY theories in $d = 4$ and $d = 6$}},}\ }\href {\doibase
  10.1007/JHEP12(2014)031} {\bibfield  {journal} {\bibinfo  {journal} {JHEP}\
  }\textbf {\bibinfo {volume} {12}},\ \bibinfo {pages} {031} (\bibinfo {year}
  {2014})},\ \Eprint {http://arxiv.org/abs/1407.6061} {arXiv:1407.6061
  [hep-th]} \BibitemShut {NoStop}%
\bibitem [{\citenamefont {Tseytlin}(2000)}]{Tseytlin:2000sf}%
  \BibitemOpen
  \bibfield  {author} {\bibinfo {author} {\bibfnamefont {Arkady~A.}\
  \bibnamefont {Tseytlin}},\ }\bibfield  {title} {\enquote {\bibinfo {title}
  {{$R^4$ terms in 11 dimensions and conformal anomaly of $(2,0)$ theory}},}\
  }\href {\doibase 10.1016/S0550-3213(00)00380-1} {\bibfield  {journal}
  {\bibinfo  {journal} {Nucl. Phys.}\ }\textbf {\bibinfo {volume} {B584}},\
  \bibinfo {pages} {233--250} (\bibinfo {year} {2000})},\ \Eprint
  {http://arxiv.org/abs/hep-th/0005072} {arXiv:hep-th/0005072 [hep-th]}
  \BibitemShut {NoStop}%
\bibitem [{\citenamefont {Mansfield}\ \emph
  {et~al.}(2003{\natexlab{b}})\citenamefont {Mansfield}, \citenamefont
  {Nolland},\ and\ \citenamefont {Ueno}}]{Mansfield:2003bg}%
  \BibitemOpen
  \bibfield  {author} {\bibinfo {author} {\bibfnamefont {Paul}\ \bibnamefont
  {Mansfield}}, \bibinfo {author} {\bibfnamefont {David}\ \bibnamefont
  {Nolland}}, \ and\ \bibinfo {author} {\bibfnamefont {Tatsuya}\ \bibnamefont
  {Ueno}},\ }\bibfield  {title} {\enquote {\bibinfo {title} {{Order $1 / N^3$
  corrections to the conformal anomaly of the $(2,0)$ theory in
  six-dimensions}},}\ }\href {\doibase 10.1016/S0370-2693(03)00777-9}
  {\bibfield  {journal} {\bibinfo  {journal} {Phys. Lett.}\ }\textbf {\bibinfo
  {volume} {B566}},\ \bibinfo {pages} {157--163} (\bibinfo {year}
  {2003}{\natexlab{b}})},\ \Eprint {http://arxiv.org/abs/hep-th/0305015}
  {arXiv:hep-th/0305015 [hep-th]} \BibitemShut {NoStop}%
\bibitem [{\citenamefont {Beccaria}\ \emph {et~al.}(2015)\citenamefont
  {Beccaria}, \citenamefont {Macorini},\ and\ \citenamefont
  {Tseytlin}}]{Beccaria:2014qea}%
  \BibitemOpen
  \bibfield  {author} {\bibinfo {author} {\bibfnamefont {Matteo}\ \bibnamefont
  {Beccaria}}, \bibinfo {author} {\bibfnamefont {Guido}\ \bibnamefont
  {Macorini}}, \ and\ \bibinfo {author} {\bibfnamefont {Arkady~A.}\
  \bibnamefont {Tseytlin}},\ }\bibfield  {title} {\enquote {\bibinfo {title}
  {{Supergravity one-loop corrections on AdS$_7$ and AdS$_3$, higher spins and
  AdS/CFT}},}\ }\href {\doibase 10.1016/j.nuclphysb.2015.01.014} {\bibfield
  {journal} {\bibinfo  {journal} {Nucl. Phys.}\ }\textbf {\bibinfo {volume}
  {B892}},\ \bibinfo {pages} {211--238} (\bibinfo {year} {2015})},\ \Eprint
  {http://arxiv.org/abs/1412.0489} {arXiv:1412.0489 [hep-th]} \BibitemShut
  {NoStop}%
\bibitem [{\citenamefont {Christensen}\ and\ \citenamefont
  {Duff}(1978)}]{Christensen:1978gi}%
  \BibitemOpen
  \bibfield  {author} {\bibinfo {author} {\bibfnamefont {S.~M.}\ \bibnamefont
  {Christensen}}\ and\ \bibinfo {author} {\bibfnamefont {M.~J.}\ \bibnamefont
  {Duff}},\ }\bibfield  {title} {\enquote {\bibinfo {title} {{Axial and
  Conformal Anomalies for Arbitrary Spin in Gravity and Supergravity}},}\
  }\href {\doibase 10.1016/0370-2693(78)90857-2} {\bibfield  {journal}
  {\bibinfo  {journal} {Phys. Lett.}\ }\textbf {\bibinfo {volume} {B76}},\
  \bibinfo {pages} {571} (\bibinfo {year} {1978})}\BibitemShut {NoStop}%
\bibitem [{\citenamefont {Christensen}\ and\ \citenamefont
  {Duff}(1979)}]{Christensen:1978md}%
  \BibitemOpen
  \bibfield  {author} {\bibinfo {author} {\bibfnamefont {S.~M.}\ \bibnamefont
  {Christensen}}\ and\ \bibinfo {author} {\bibfnamefont {M.~J.}\ \bibnamefont
  {Duff}},\ }\bibfield  {title} {\enquote {\bibinfo {title} {{New Gravitational
  Index Theorems and Supertheorems}},}\ }\href {\doibase
  10.1016/0550-3213(79)90516-9} {\bibfield  {journal} {\bibinfo  {journal}
  {Nucl. Phys.}\ }\textbf {\bibinfo {volume} {B154}},\ \bibinfo {pages}
  {301--342} (\bibinfo {year} {1979})}\BibitemShut {NoStop}%
\bibitem [{\citenamefont {Gilkey}(1975)}]{Gilkey:1975iq}%
  \BibitemOpen
  \bibfield  {author} {\bibinfo {author} {\bibfnamefont {Peter~B.}\
  \bibnamefont {Gilkey}},\ }\bibfield  {title} {\enquote {\bibinfo {title}
  {{The Spectral geometry of a Riemannian manifold}},}\ }\href {\doibase
  10.4310/jdg/1214433164} {\bibfield  {journal} {\bibinfo  {journal} {J. Diff.
  Geom.}\ }\textbf {\bibinfo {volume} {10}},\ \bibinfo {pages} {601--618}
  (\bibinfo {year} {1975})}\BibitemShut {NoStop}%
\bibitem [{\citenamefont {Bastianelli}\ \emph {et~al.}(2000)\citenamefont
  {Bastianelli}, \citenamefont {Frolov},\ and\ \citenamefont
  {Tseytlin}}]{Bastianelli:2000hi}%
  \BibitemOpen
  \bibfield  {author} {\bibinfo {author} {\bibfnamefont {F.}~\bibnamefont
  {Bastianelli}}, \bibinfo {author} {\bibfnamefont {S.}~\bibnamefont {Frolov}},
  \ and\ \bibinfo {author} {\bibfnamefont {Arkady~A.}\ \bibnamefont
  {Tseytlin}},\ }\bibfield  {title} {\enquote {\bibinfo {title} {{Conformal
  anomaly of $(2,0)$ tensor multiplet in six-dimensions and AdS / CFT
  correspondence}},}\ }\href {\doibase 10.1088/1126-6708/2000/02/013}
  {\bibfield  {journal} {\bibinfo  {journal} {JHEP}\ }\textbf {\bibinfo
  {volume} {02}},\ \bibinfo {pages} {013} (\bibinfo {year} {2000})},\ \Eprint
  {http://arxiv.org/abs/hep-th/0001041} {arXiv:hep-th/0001041 [hep-th]}
  \BibitemShut {NoStop}%
\bibitem [{\citenamefont {Parker}\ and\ \citenamefont
  {Rosenberg}(1987)}]{parker1987}%
  \BibitemOpen
  \bibfield  {author} {\bibinfo {author} {\bibfnamefont {Thomas}\ \bibnamefont
  {Parker}}\ and\ \bibinfo {author} {\bibfnamefont {Steven}\ \bibnamefont
  {Rosenberg}},\ }\bibfield  {title} {\enquote {\bibinfo {title} {Invariants of
  conformal laplacians},}\ }\href {\doibase 10.4310/jdg/1214440850} {\bibfield
  {journal} {\bibinfo  {journal} {J. Diff. Geom.}\ }\textbf {\bibinfo {volume}
  {25}},\ \bibinfo {pages} {199--222} (\bibinfo {year} {1987})}\BibitemShut
  {NoStop}%
\bibitem [{\citenamefont {Minwalla}(1998)}]{Minwalla:1997ka}%
  \BibitemOpen
  \bibfield  {author} {\bibinfo {author} {\bibfnamefont {Shiraz}\ \bibnamefont
  {Minwalla}},\ }\bibfield  {title} {\enquote {\bibinfo {title} {{Restrictions
  imposed by superconformal invariance on quantum field theories}},}\
  }\href@noop {} {\bibfield  {journal} {\bibinfo  {journal} {Adv. Theor. Math.
  Phys.}\ }\textbf {\bibinfo {volume} {2}},\ \bibinfo {pages} {781--846}
  (\bibinfo {year} {1998})},\ \Eprint {http://arxiv.org/abs/hep-th/9712074}
  {arXiv:hep-th/9712074 [hep-th]} \BibitemShut {NoStop}%
\bibitem [{\citenamefont {Dobrev}(2002)}]{Dobrev:2002dt}%
  \BibitemOpen
  \bibfield  {author} {\bibinfo {author} {\bibfnamefont {V.~K.}\ \bibnamefont
  {Dobrev}},\ }\bibfield  {title} {\enquote {\bibinfo {title} {{Positive energy
  unitary irreducible representations of$D = 6$ conformal supersymmetry}},}\
  }\href {\doibase 10.1088/0305-4470/35/33/308} {\bibfield  {journal} {\bibinfo
   {journal} {J. Phys.}\ }\textbf {\bibinfo {volume} {A35}},\ \bibinfo {pages}
  {7079--7100} (\bibinfo {year} {2002})},\ \Eprint
  {http://arxiv.org/abs/hep-th/0201076} {arXiv:hep-th/0201076 [hep-th]}
  \BibitemShut {NoStop}%
\bibitem [{\citenamefont {Bhattacharya}\ \emph {et~al.}(2008)\citenamefont
  {Bhattacharya}, \citenamefont {Bhattacharyya}, \citenamefont {Minwalla},\
  and\ \citenamefont {Raju}}]{Bhattacharya:2008zy}%
  \BibitemOpen
  \bibfield  {author} {\bibinfo {author} {\bibfnamefont {Jyotirmoy}\
  \bibnamefont {Bhattacharya}}, \bibinfo {author} {\bibfnamefont {Sayantani}\
  \bibnamefont {Bhattacharyya}}, \bibinfo {author} {\bibfnamefont {Shiraz}\
  \bibnamefont {Minwalla}}, \ and\ \bibinfo {author} {\bibfnamefont {Suvrat}\
  \bibnamefont {Raju}},\ }\bibfield  {title} {\enquote {\bibinfo {title}
  {{Indices for Superconformal Field Theories in 3, 5 and 6 Dimensions}},}\
  }\href {\doibase 10.1088/1126-6708/2008/02/064} {\bibfield  {journal}
  {\bibinfo  {journal} {JHEP}\ }\textbf {\bibinfo {volume} {02}},\ \bibinfo
  {pages} {064} (\bibinfo {year} {2008})},\ \Eprint
  {http://arxiv.org/abs/0801.1435} {arXiv:0801.1435 [hep-th]} \BibitemShut
  {NoStop}%
\bibitem [{\citenamefont {Buican}\ \emph {et~al.}(2016)\citenamefont {Buican},
  \citenamefont {Hayling},\ and\ \citenamefont
  {Papageorgakis}}]{Buican:2016hpb}%
  \BibitemOpen
  \bibfield  {author} {\bibinfo {author} {\bibfnamefont {Matthew}\ \bibnamefont
  {Buican}}, \bibinfo {author} {\bibfnamefont {Joseph}\ \bibnamefont
  {Hayling}}, \ and\ \bibinfo {author} {\bibfnamefont {Constantinos}\
  \bibnamefont {Papageorgakis}},\ }\bibfield  {title} {\enquote {\bibinfo
  {title} {{Aspects of Superconformal Multiplets in $D>4$}},}\ }\href {\doibase
  10.1007/JHEP11(2016)091} {\bibfield  {journal} {\bibinfo  {journal} {JHEP}\
  }\textbf {\bibinfo {volume} {11}},\ \bibinfo {pages} {091} (\bibinfo {year}
  {2016})},\ \Eprint {http://arxiv.org/abs/1606.00810} {arXiv:1606.00810
  [hep-th]} \BibitemShut {NoStop}%
\bibitem [{\citenamefont {Cordova}\ \emph
  {et~al.}(2016{\natexlab{a}})\citenamefont {Cordova}, \citenamefont
  {Dumitrescu},\ and\ \citenamefont {Intriligator}}]{Cordova:2016emh}%
  \BibitemOpen
  \bibfield  {author} {\bibinfo {author} {\bibfnamefont {Clay}\ \bibnamefont
  {Cordova}}, \bibinfo {author} {\bibfnamefont {Thomas~T.}\ \bibnamefont
  {Dumitrescu}}, \ and\ \bibinfo {author} {\bibfnamefont {Kenneth}\
  \bibnamefont {Intriligator}},\ }\bibfield  {title} {\enquote {\bibinfo
  {title} {{Multiplets of Superconformal Symmetry in Diverse Dimensions}},}\
  }\href@noop {} {\  (\bibinfo {year} {2016}{\natexlab{a}})},\ \Eprint
  {http://arxiv.org/abs/1612.00809} {arXiv:1612.00809 [hep-th]} \BibitemShut
  {NoStop}%
\bibitem [{\citenamefont {Cordova}\ \emph
  {et~al.}(2016{\natexlab{b}})\citenamefont {Cordova}, \citenamefont
  {Dumitrescu},\ and\ \citenamefont {Intriligator}}]{Cordova:2015fha}%
  \BibitemOpen
  \bibfield  {author} {\bibinfo {author} {\bibfnamefont {Clay}\ \bibnamefont
  {Cordova}}, \bibinfo {author} {\bibfnamefont {Thomas~T.}\ \bibnamefont
  {Dumitrescu}}, \ and\ \bibinfo {author} {\bibfnamefont {Kenneth}\
  \bibnamefont {Intriligator}},\ }\bibfield  {title} {\enquote {\bibinfo
  {title} {{Anomalies, renormalization group flows, and the a-theorem in
  six-dimensional $(1, 0)$ theories}},}\ }\href {\doibase
  10.1007/JHEP10(2016)080} {\bibfield  {journal} {\bibinfo  {journal} {JHEP}\
  }\textbf {\bibinfo {volume} {10}},\ \bibinfo {pages} {080} (\bibinfo {year}
  {2016}{\natexlab{b}})},\ \Eprint {http://arxiv.org/abs/1506.03807}
  {arXiv:1506.03807 [hep-th]} \BibitemShut {NoStop}%
\bibitem [{\citenamefont {Beccaria}\ and\ \citenamefont
  {Tseytlin}(2016)}]{Beccaria:2015ypa}%
  \BibitemOpen
  \bibfield  {author} {\bibinfo {author} {\bibfnamefont {Matteo}\ \bibnamefont
  {Beccaria}}\ and\ \bibinfo {author} {\bibfnamefont {Arkady~A.}\ \bibnamefont
  {Tseytlin}},\ }\bibfield  {title} {\enquote {\bibinfo {title} {{Conformal
  anomaly c-coefficients of superconformal 6d theories}},}\ }\href {\doibase
  10.1007/JHEP01(2016)001} {\bibfield  {journal} {\bibinfo  {journal} {JHEP}\
  }\textbf {\bibinfo {volume} {01}},\ \bibinfo {pages} {001} (\bibinfo {year}
  {2016})},\ \Eprint {http://arxiv.org/abs/1510.02685} {arXiv:1510.02685
  [hep-th]} \BibitemShut {NoStop}%
\bibitem [{\citenamefont {Yankielowicz}\ and\ \citenamefont
  {Zhou}(2017)}]{Yankielowicz:2017xkf}%
  \BibitemOpen
  \bibfield  {author} {\bibinfo {author} {\bibfnamefont {Shimon}\ \bibnamefont
  {Yankielowicz}}\ and\ \bibinfo {author} {\bibfnamefont {Yang}\ \bibnamefont
  {Zhou}},\ }\bibfield  {title} {\enquote {\bibinfo {title} {{Supersymmetric
  Renyi entropy and Anomalies in 6d $(1,0)$ SCFTs}},}\ }\href {\doibase
  10.1007/JHEP04(2017)128} {\bibfield  {journal} {\bibinfo  {journal} {JHEP}\
  }\textbf {\bibinfo {volume} {04}},\ \bibinfo {pages} {128} (\bibinfo {year}
  {2017})},\ \Eprint {http://arxiv.org/abs/1702.03518} {arXiv:1702.03518
  [hep-th]} \BibitemShut {NoStop}%
\bibitem [{\citenamefont {Beccaria}\ and\ \citenamefont
  {Tseytlin}(2017)}]{Beccaria:2017dmw}%
  \BibitemOpen
  \bibfield  {author} {\bibinfo {author} {\bibfnamefont {M.}~\bibnamefont
  {Beccaria}}\ and\ \bibinfo {author} {\bibfnamefont {A.~A.}\ \bibnamefont
  {Tseytlin}},\ }\bibfield  {title} {\enquote {\bibinfo {title} {{C$_{T}$ for
  higher derivative conformal fields and anomalies of $(1, 0)$ superconformal
  6d theories}},}\ }\href {\doibase 10.1007/JHEP06(2017)002} {\bibfield
  {journal} {\bibinfo  {journal} {JHEP}\ }\textbf {\bibinfo {volume} {06}},\
  \bibinfo {pages} {002} (\bibinfo {year} {2017})},\ \Eprint
  {http://arxiv.org/abs/1705.00305} {arXiv:1705.00305 [hep-th]} \BibitemShut
  {NoStop}%
\bibitem [{\citenamefont {Lichnerowicz}\ and\ \citenamefont
  {Møller}(1962)}]{Lich}%
  \BibitemOpen
  \bibfield  {author} {\bibinfo {author} {\bibfnamefont {A.}~\bibnamefont
  {Lichnerowicz}}\ and\ \bibinfo {author} {\bibfnamefont {C.}~\bibnamefont
  {Møller}},\ }\bibfield  {title} {\enquote {\bibinfo {title} {Propagators and
  commutators in general relativity [and discussion]},}\ }\href
  {http://www.jstor.org/stable/2414535} {\bibfield  {journal} {\bibinfo
  {journal} {Proceedings of the Royal Society of London. Series A, Mathematical
  and Physical Sciences}\ }\textbf {\bibinfo {volume} {270}},\ \bibinfo {pages}
  {342--345} (\bibinfo {year} {1962})}\BibitemShut {NoStop}%
\bibitem [{\citenamefont {van Ritbergen}\ \emph {et~al.}(1999)\citenamefont
  {van Ritbergen}, \citenamefont {Schellekens},\ and\ \citenamefont
  {Vermaseren}}]{GroupTheory}%
  \BibitemOpen
  \bibfield  {author} {\bibinfo {author} {\bibfnamefont {T.}~\bibnamefont {van
  Ritbergen}}, \bibinfo {author} {\bibfnamefont {A.~N.}\ \bibnamefont
  {Schellekens}}, \ and\ \bibinfo {author} {\bibfnamefont {J.~A.~M.}\
  \bibnamefont {Vermaseren}},\ }\bibfield  {title} {\enquote {\bibinfo {title}
  {{Group theory factors for Feynman diagrams}},}\ }\href {\doibase
  10.1142/S0217751X99000038} {\bibfield  {journal} {\bibinfo  {journal} {Int.
  J. Mod. Phys.}\ }\textbf {\bibinfo {volume} {A14}},\ \bibinfo {pages}
  {41--96} (\bibinfo {year} {1999})},\ \Eprint
  {http://arxiv.org/abs/hep-ph/9802376} {arXiv:hep-ph/9802376 [hep-ph]}
  \BibitemShut {NoStop}%
\bibitem [{\citenamefont {Okubo}(1982)}]{Okubo:1981td}%
  \BibitemOpen
  \bibfield  {author} {\bibinfo {author} {\bibfnamefont {Susumu}\ \bibnamefont
  {Okubo}},\ }\bibfield  {title} {\enquote {\bibinfo {title} {{Modified Fourth
  Order Casimir Invariants and Indices for Simple Lie Algebras}},}\ }\href
  {\doibase 10.1063/1.525212} {\bibfield  {journal} {\bibinfo  {journal} {J.
  Math. Phys.}\ }\textbf {\bibinfo {volume} {23}},\ \bibinfo {pages} {8}
  (\bibinfo {year} {1982})}\BibitemShut {NoStop}%
\end{thebibliography}%
\end{document}